\documentclass[11pt, dvips]{elsart}
\usepackage{graphicx}
\usepackage{epsfig}
\begin{document}
\bibliographystyle{unsrt}

\journal{Astroparticle Physics}

%%%%%%%%%%%%%%%%%%%%%%%%%%%%%%%%%%%%%%%%%%%%%%%%%%%%%%%%%%%%%%%%%%%%%%%%%%%%%%%
% BLANCA special commands and standards
%\hyphenation{Cher-en-kov}
\hyphenation{A-MAN-DA} \hyphenation{VUL-CAN}
\newcommand{\unit}[1]{\ensuremath{\,\mathrm{#1}}}
\newcommand{\gcm}[1]{\ensuremath{#1\unit{g\,cm}^{-2}}}
\renewcommand{\etal}{\emph{et al.}}
\newcommand{\Xmax}{\ensuremath{X_{max}}}
\newcommand{\meanXmax}{\ensuremath{\langle X_{max} \rangle}}
\newcommand{\lnA}{\ensuremath{\ln(A)}}
\newcommand{\meanlnA}{\ensuremath{\langle \ln(A) \rangle}}
\newcommand{\exposure}{\ensuremath{1.83\times10^{10} \unit{m}^2 \unit{sr}
\unit{s}}}

%%%%%%%%%%%%%%%%%%%%%%%%%%%%%%%%%%%%%%%%%%%%%%%%%%%%%%%%%%%%%%%%%%%%%%%%%%%%%%%
\begin{frontmatter}
\title{The Composition of Cosmic Rays at the Knee}

\author[EFI]{S.P. Swordy}
\author[Adler,EFI]{L.F. Fortson}
\author[EFI]{J. Hinton}
\author[EFI,karl]{J. H\"orandel}
\author[Leeds]{J. Knapp}
\author[EFI]{C.L. Pryke}
\author[aoy]{T. Shibata}
\author[EFI]{S.P. Wakely}
\author[Utah]{Z. Cao}
\author[LSU]{M. L. Cherry}
\author[PSU]{S. Coutu}
\author[EFI]{J. Cronin}
\author[Bart]{R. Engel}
\author[EFI,Princeton]{J.W. Fowler}
\author[karl]{K.- H. Kampert}
\author[mpi]{J. Kettler}
\author[Utah]{D.B. Kieda}
\author[LSU]{J. Matthews}
\author[PSU]{S. A. Minnick}
\author[GSFC]{A. Moiseev}
\author[EFI]{D. Muller}
\author[karl]{M. Roth}
\author[TTU]{A. Sill}
\author[Bart]{G. Spiczak}

\address[EFI]{The Enrico Fermi Institute, University of Chicago,
              5640 Ellis Avenue, Chicago, Illinois 60637-1433, USA}
\address[Adler]{Dept.\ of Astronomy, Adler Planetarium and Astronomy
Museum, Chicago, Illinois 60605, USA}
\address[Leeds]{Dept.\ of Physics and Astronomy, University of Leeds,
    Leeds LS2 9JT, U.K.}
\address[aoy]{Dept.\ of Physics, Aoyama Gakuin University, Tokyo, 157-8572,
    Japan}
\address[Utah]{High Energy Astrophysics Institute, Dept.\ of Physics,
University of Utah, Salt Lake City, Utah 84112, USA}
\address[LSU]{Dept.\ of Physics and Astronomy, Louisiana State University,
             Baton Rouge, LA 70803, USA}
\address[PSU]{Dept.\ of Physics, Penn State University, University Park,
          PA 16802, USA}
\address[Bart]{Bartol Research Institute, University of Delaware, Newark,
    DE 19716, USA}
\address[Princeton]{Department of Physics, Princeton University, Princeton,
   NJ 08544, USA}
\address[karl]{Institut f\"{u}r Experimentelle Kernphysik, University of
    Karlsruhe, D-76021 Karlsruhe, Germany}
\address[mpi]{Max-Planck-Institut f\"{u}r Kernphysik, Saupfercheckweg 1,
    D-69117 Heidelberg, Germany}
\address[GSFC]{NASA, GSFC, Code 660, Greenbelt, MD 20771, USA}
\address[TTU] {Dept. of Physics, Texas Tech University, Lubbock, TX 79409, USA
}

%\corauth[CA]{Corresponding author.}
%: Enrico Fermi Institute
%              5640 S. Ellis Ave, University of Chicago, Chicago, IL
%              60637, USA.  Email: swordy@ulysses.uchicago.edu Phone:
%              1-773-702-7835.}

%%%%%%%%%%%%%%%%%%%%%%%%%%%%%%%%%%%%%%%%%%%%%%%%%%%%%%%%%%%%%%%%%%%%%%%%%%%%%%%
\begin{abstract}
The observation of a small change in spectral slope, or 'knee' in
the fluxes of cosmic rays near energies $10^{15}$eV has caused
much speculation since its discovery over 40 years ago. The origin
of this feature remains unknown. A small workshop to review some
modern experimental measurements of this region was held at the
Adler Planetarium in Chicago, USA in June 2000. This paper
summarizes the results presented at this workshop and the
discussion of their interpretation in the context of hadronic
models of atmospheric airshowers.
\end{abstract}

\end{frontmatter}

%%%%%%%%%%%%%%%%%%%%%%%%%%%%%%%%%%%%%%%%%%%%%%%%%%%%%%%%%%%%%%%%%%%%%%%%%%%%%%%
\section{Introduction} \label{sec.intro}
The primary cosmic ray particles extend over at least twelve
decades of energy with a corresponding decline in intensity of
over thirty decades. The spectrum is remarkably featureless with
little deviation from a constant power law across this large
energy range. The small change in slope, from $\propto E^{-2.7}$
to $\propto E^{-3.0}$, near $10^{15}$~eV is known as the `knee' of
the spectrum. The coincidence of this feature with the highest
energies expected from diffusive shock acceleration in supernova
remnants is intriguing~\cite{ces83}. It can also be attributed to
contributions from nearby or recent supernova events~\cite{erl97}.
To distinguish between these various ideas, better measurements of
the elemental composition of cosmic rays at these energies are
essential. The accurate determination of composition has provided
some of the key advances in this field at lower energies, where
direct measurements are possible with detectors above the
atmosphere. For example, the realization that the observed
spectral slope of cosmic rays is significantly steeper than that
produced in the cosmic ray sources themselves resulted from
measurements with sufficient elemental resolution to separate
primary source cosmic ray elements from those produced in the
interstellar medium at $10^{11}$ eV~\cite{jul72}.

Here we collect together various recent measurements of cosmic
rays in the `knee' region which were discussed at a workshop held
in Chicago in June 2000 (http://knee.uchicago.edu). The objective
of this paper is to provide an overview of the state of
experimental measurements in this field and to provide some
suggestions for future improvements. The systematic limitations of
the type of measurements presented here are also evaluated. In the
modern era, these limitations are associated in part with the
validity of the various numerical simulations used to interpret
the data and the absolute energy scale for the data.

The structure of this paper basically keeps to the structure of
the workshop where both experimental results and various aspects
of the numerical simulations were discussed. This work is not an
exhaustive summary of the field but summarizes the efforts that
were represented at the workshop. There are other notable
experiments in this area which were not discussed in the workshop,
in particular HEGRA\cite{arq00} and EAS-TOP\cite{agl99}. Also,
developments since the workshop are not discussed here, these are
summarized in the proceedings of the 27th International Cosmic Ray
Conference (2001) Hamburg\cite{som01}.

We begin with a brief overview of the relative merits of the
experimental techniques followed by short descriptions of direct
and indirect (air shower) measurements presented at the meeting.
There follows a relatively detailed discussion of air shower
simulations and the physics which goes into them. We then present
a simple comparison of the results of the experiments in terms of
the mean logarithm of the inferred primary mass. Contemporary
experiments are breaking new ground in terms of statistics and
modeling, and we include some highlights of multi-species fits and
multi-parameter correlation studies. Finally we draw some
conclusions.

\begin{figure}
  \centering
  \includegraphics[height=4.0in]{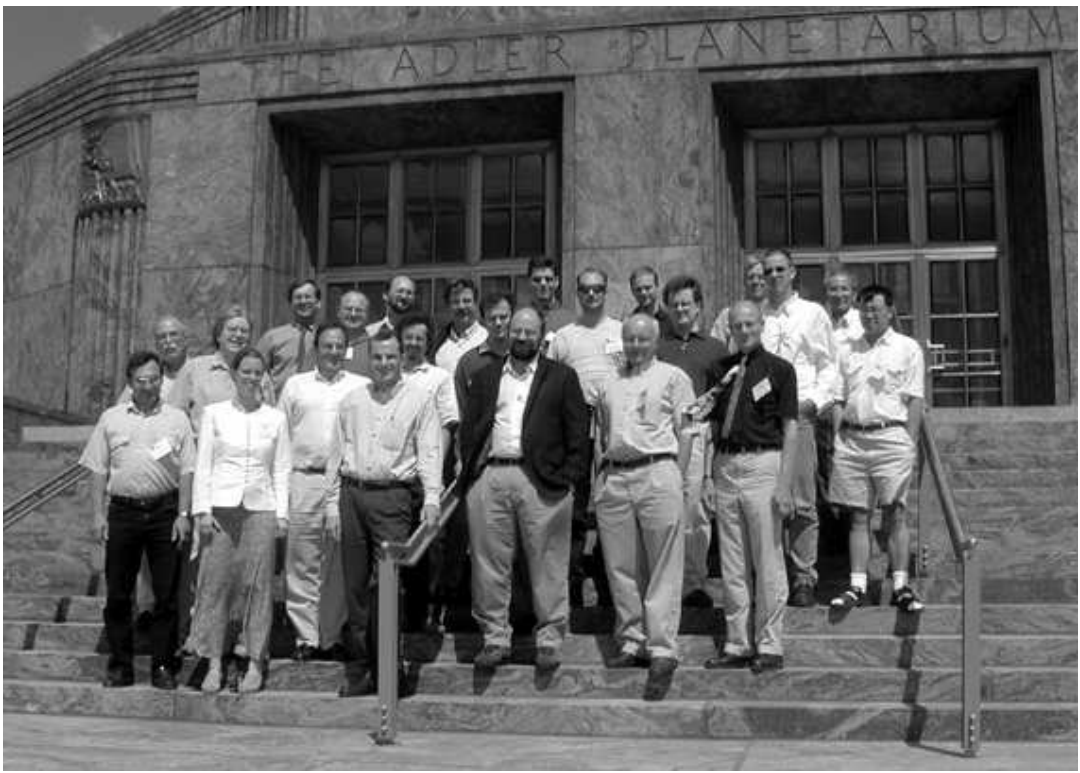}
  \caption{Participants in the cosmic ray workshop at Adler Planetarium. (Left to Right)
  Moissev, Matthews, Swordy, Fortson, Sills, Cherry, Minnick, H\"orandel, Kampert, Coutu,
  Knapp, Roth, Kieda, Wakely, Pryke, M\"uller, Hinton, Fowler, Cronin, Spiczak, Kettler,
  Shibata, Cao }
  \label{direct}
\end{figure}
%%%%%%%%%%%%%%%%%%%%%%%%%%%%%%%%%%%%%%%%%%%%%%%%%%%%%%%%%%%%%%%%%%%%%%%%%%%%%%%
\section{Overview of Techniques}\label{sec.tech}
This section seeks to summarize the main cosmic ray detection
techniques and their relative merits. By no means comprehensive,
the points that are touched on here were felt to be most relevant
to the experiments discussed at the meeting. The experiment
acronyms are defined in Section~\ref{sec.exp}.

%%%%%%%%%%%%%%%%%%%%%%%%%%%%%%%%%%%%%%
\subsection{Direct Detection}

The detection of cosmic rays above the atmosphere is the only way
to obtain direct measurements of the primary particles and their
energy spectra. There are a number of techniques for direct
detection. So-called ``active'' techniques usually involve some
combination of charge detection instruments and energy measurement
instruments,  such as scintillator combined with a  calorimeter or
transition radiation detector (TRD). Here, the data can be
recorded electronically in-flight. TRD's  rely on the passage of a
particle through the detector without a nuclear interaction while
calorimeters not only require an interaction but must contain as
much of the ensuing cascade as possible for accurate energy
measurements. It should be noted that present TRD's saturate at
about $10^{15}$ eV and provide better energy measurements of high
Z nuclei compared to protons and Helium. In contrast, calorimetric
techniques can be used to obtain reasonable energy measurements of
protons and Helium up to knee energies.

Nuclear emulsion is an example of a passive technique that, when
interleaved with suitable converters, can simultaneously measure
the primary charge and energy. These payloads must be recovered
and subjected to complicated and exacting procedures to extract
the energy and charge information. While each technique has its
own strengths and weaknesses, the major limitation to all direct
techniques to date is the inability to fly large area detectors
for long periods of time. This limit is constantly being pushed as
with the recent advent of Ultra-Long-Duration-Balloon flights and
by constructing payloads that do not require pressurized vessels
for operation. However, the nuclear emulsion experiments currently
have the largest exposure and therefore the most statistically
significant data at the highest energies. It is these experiments
that will be discussed below.

%%%%%%%%%%%%%%%%%%%%%%%%%%%%%%%%%%%%%%
\subsection{Indirect Detection}

Once the primary particle interacts with the Earth's atmosphere,
the detection techniques indirectly measure the primary parameters
through the air shower components produced in the interaction. It
is then a matter of reconstructing the primary energy and nuclear
charge from the air shower observables. The main difficulty with
indirect techniques is that these reconstructions depend on
hadronic interactions which are empirically undetermined at this
time for the relevant energies and kinematical regions. However,
large area, long duration experiments can be operated from the
ground allowing statistically significant measurements to be made
up to the highest energies in the cosmic ray spectrum.

The different techniques can be separated into detectors that
measure the particle content of the shower (KASCADE, CASA-MIA),
and those that measure Cherenkov light (DICE, BLANCA, VULCAN).

The slope and normalization of the lateral distribution of an air
shower component are functions of the energy and mass of the
primary cosmic ray. While composition information can be obtained
by lateral distribution measurements from a charged particle array
(e.g.\ the CASA-MIA results below use the electron lateral
distribution slope), in practice, it is much easier to make
density measurements of the required precision when working with
Cherenkov light. Because many photons are emitted by a given
charged particle, and these photons are not significantly
attenuated by the atmosphere, they are much more numerous than
charged particles at ground level. Additionally, since the
majority of the light is emitted well above the ground, and at
significant angles to the shower axis, its lateral distribution at
the ground is flatter than that of charged particles. For these
reasons, the sensitivity and dynamic range requirements of
Cherenkov detectors are not as stringent as those for charged
particle detectors to probe the primary composition in a given
energy range. These are the strengths behind the BLANCA and VULCAN
results discussed below. However, this advantage is at least
partially offset by the requirement that Cherenkov detectors can
only run on dark, moonless nights.

The DICE experiment measures the angular distribution of Cherenkov
light, and thus makes a somewhat direct measurement of the
longitudinal profile of the air shower. This puts it a different
class to all of the other experiments discussed here\footnote
{With the exception of the KASCADE low energy muon tracking
detectors which are not mentioned in this paper.} which measure
quantities which have been ``integrated'' over the shower
development history.

The multi-component nature of air showers allows event-by-event
cross-correlation measurements. This approach has been widely
exploited by the KASCADE experiment, it is also possible with the
DICE and BLANCA experiments which are co-located with CASA-MIA
(see Section~\ref{sec.multi}).

%%%%%%%%%%%%%%%%%%%%%%%%%%%%%%%%%%%%%%%%%%%%%%%%%%%%%%%%%%%%%%%%%%%%%%%%%%%%%%%
\section{Overview of Experiments}\label{sec.exp}

Here we provide a brief description of the measurements discussed
at the meeting, including direct detections by RUNJOB and JACEE
and the air shower experiments KASCADE, CASA-MIA, DICE, BLANCA and
SPASE-VULCAN.

%%%%%%%%%%%%%%%%%%%%%%%%%%%%%%%%%%%%%%
\subsection{JACEE} \label{sec.jacee}

The Japanese-American Cooperative Emulsion Experiment (JACEE) is a
series of high altitude balloon flights for direct detection of
cosmic rays. The technique uses an initial charge-detecting layer
for identifying the elemental charge, a target section for
providing a nuclear interaction, and a calorimeter layer which
collects the fraction of incident particle energy which appears as
gamma-rays in the interaction ($\Sigma E_\gamma$).  In order to
accumulate the required high energy statistics, JACEE has now
flown emulsion chambers on 15 balloon flights (eight 1-2 day
turnaround flights, two 5-6 day Australia-to-South America
flights, and five 9-15 day Antarctic circumpolar flights
~\cite{wil95}). All but one of these have been successfully
recovered. The total accumulated exposure is 1436 m$^2$~hrs. The
average flight altitude ranges from 3.5 to 5.5 g cm$^2$. A single
flight typically carries 2-6 emulsion blocks, each generally
40$\times$50 cm$^2$. Fifty-eight emulsion blocks have been flown ;
fifty-two have been recovered. The data discussed here come from
JACEE flights 1-12, covering the results from the first 40
emulsion blocks (cumulative exposure 644 m$^2$hrs). The total
number of high energy events available for analysis from all
flights is $\sim 2\times 10^4$. Of these, $\sim 180$ have energy
exceeding 100 TeV per particle. The present analysis is based on
656 protons above 6 TeV and 414 helium nuclei above a total energy
of 8 TeV per particle. The basic detector used in the JACEE
experiments is a fine grained emulsion~\cite{bur86} chamber
typically containing approximately a hundred track­ sensitive
nuclear emulsion plates and a three dimensional
emulsion/x-ray/lead plate calorimeter. The lower part of the
chamber is the calorimeter section, consisting of $\sim$20 layers
of emulsions and x­ray films interleaved with up to 8.5 radiation
lengths of 1-2.5 mm thick lead plates. The calorimeter records
singly charged particle tracks with a spatial resolution in the
emulsion of better than 1 $\mu$m and individual photon cascades
with a resolution of a few microns. High energy showers produce
visible dark spots in the x­ray film, which are used to locate and
trace the energetic cascades. On average, more than 400 events are
detected per block with an optical density corresponding to a
total energy in the electromagnetic shower $\Sigma E_\gamma \ge
1.5$TeV for protons. In the original JACEE
analyses~\cite{bur86,bur90,asa91,asa93} electron counts in the
emulsion layers along the cascade were compared to a simulated
shower development curve to determine the total electromagnetic
energy deposited in the calorimeter. In order to speed up the
analysis for the large new data sample obtained from the long
duration Antarctic flights, for these flights the energy is
derived by fitting the observed darkness measured in the x-ray
films to electromagnetic shower development curves to determine
$\Sigma E_\gamma$~\cite{asa93,ols95}. The results of the measured
fluxes of protons and helium nuclei into the knee region are shown
in Figure~\ref{direct}. A fit to the unbinned integral spectra
gives a proton spectrum with a differential power law index $2.80
\pm 0.04$ and a helium spectrum with a differential index $2.68 +
0.04 / -0.06$. There is no evidence of a break in either spectrum
out to at least 40-90 TeV\cite{cherry99}.

\begin{figure}
  \centering
  \includegraphics[height=4.0in]{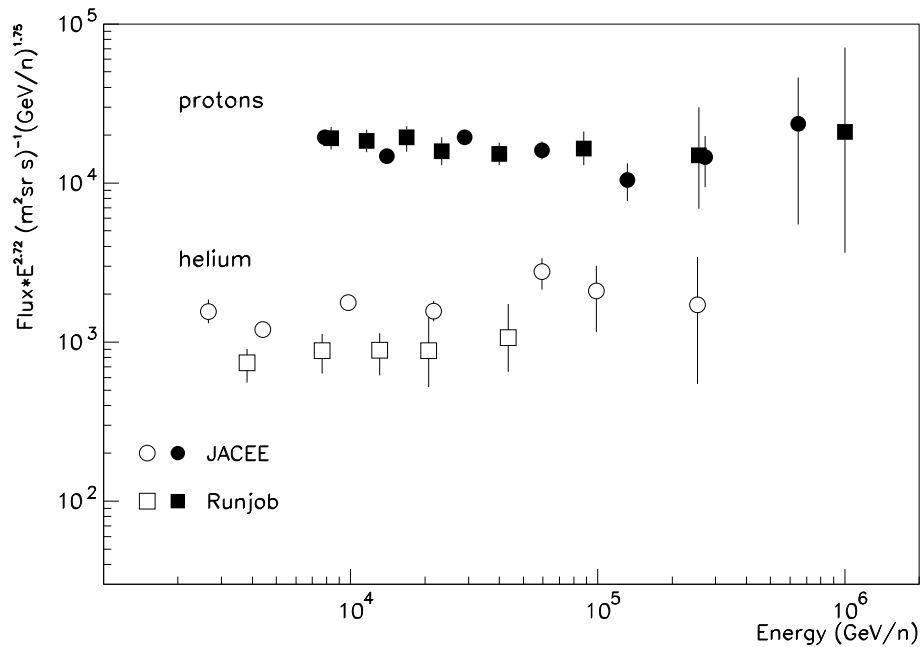}
  \caption{Direct measurements of cosmic ray components by the
           JACEE~\cite{asa98} and RUNJOB~\cite{apa99} experiments}
  \label{direct}
\end{figure}

%%%%%%%%%%%%%%%%%%%%%%%%%%%%%%%%%%%%%%
\subsection{RUNJOB} \label{sec.runjob}

The RUssian-Nippon Joint Balloon collaboration (RUNJOB) has been
carrying out Trans-Siberian continental balloon flights since
1995, and eleven balloons have been launched from the Kamchatka
peninsula, of which ten were successful. The payloads directly
detect cosmic rays using a stack of passive detectors. Twenty
emulsion chambers in total, each having an area of
40~cm$\times$50~cm, have been recovered near the Volga region or
west side of the Ural Mountains after an exposure of approximately
150 hours. The total exposure factor amounts to 575 m$^2$ hrs. In
this work the results based on the full analysis of the '95 and
'96 experiments and one quarter of the '97 experiment are given.
The exposure factor used for the present data analysis amounts to
261.5 m$^2$ hr. The performance of the balloon campaign and the
data processing are presented elsewhere~\cite{apa01}.

The payload mass of these flights is limited to 250 kg in each
flight to ensure an altitude above $\sim$10 g/cm$^2$. The detector
chambers consist of 4 parts: the Primary layer, Target layer,
Spacer layer, and Calorimeter layer (see~\cite{apa99}). The
primary layer is an emulsion used for charge identification, the
calorimeter consists of emulsion and x-ray films. The thickness of
each layer is optimized based on the experience of our analysis.
There is a wide spacer between target module (made of lucite or
stainless steel plate) and the calorimeter module, which allows a
measurement of the opening angle of secondary $\gamma$-rays
($\pi_0 \rightarrow 2\gamma$) produced by the nuclear interaction
at the target. The spacer thickness is typically 15-20 cm in the
shower development direction, which gives a measurement of the
opening angle for individual high energy $\gamma$-rays with energy
20-50 TeV, taking the inclination effect into account. An estimate
can be made of the total shower energy with $E_\gamma\sim$100~TeV
or more, where the electromagnetic shower maximum may not be
contained in the thin calorimeter, although this does depend on
the inclination of the shower. The fraction of the incoming
particle energy which is released into the $\gamma$-ray component
can be estimated by using cascade shower theory, which is
well-established through accelerator experiments~\cite{hot80}. The
maximum darkness of the shower in the x-ray film is approximately
proportional to the shower energy, $\Sigma E_\gamma$, independent
of the primary particle mass. The spot darkness is measured with
use of a photo-densitometer, the details of which are summarized
in~\cite{oka87} and~\cite{fuj89}. However, the vertical thickness
of the RUNJOB chamber is as thin as 4-5 radiation lengths. A
method to overcome this limitation has been discussed
previously~\cite{apa99a}. For heavy nuclei, the opening-angle
method for fragments can be used. This method is summarized in
detail in~\cite{ich93}. The results of the measured fluxes of
protons and helium into the knee region are shown in
Figure~\ref{direct}.

%%%%%%%%%%%%%%%%%%%%%%%%%%%%%%%%%%%%%%
\subsection{KASCADE} \label{sec.kascade}

The KASCADE (KArlsruhe Shower Core and Array DEtector) array is an
air shower detector which consists of 252 detector stations in a
$200\times 200$m$^2$ rectangular grid located near Karlsruhe,
Germany~\cite{kla97}. The array comprises unshielded liquid
scintillation detectors ($e/\gamma$ detectors) and plastic
scintillators as muon detectors below 10 cm of lead and 4 cm of
steel. The total sensitive areas are 490 m$^2$ for the $e/\gamma$
detectors and 622\ m$^2$ for the muon detectors. In the center of
the array a hadron calorimeter ($16\times 20$\ m$^2$ ) is built
up, consisting of more than 40,000 liquid ionization chambers in 8
layers~\cite{eng99}. In between the calorimeter a trigger layer
consisting of 456 scintillation detectors of 0.456 m$^2$ each,
measures the energy deposit and arrival times of muons with
E$>$490 MeV. Below the calorimeter an array of position sensitive
multiwire proportional chambers in two layers measures EAS muons
with $E_\mu >2.4$GeV. Various methods and results of the analysis
of KASCADE data are described in detail in the proceedings of the
last ICRC conference in Salt Lake City, Utah~\cite{ant99} or
in~\cite{kam99}, where the KASCADE contributions are sampled. The
main observables of KASCADE per single shower are the so called
shower sizes, i.e. total number of electrons $N_e$ and number of
muons within the range of the core distance 40-200m, $N^{tr}_\mu$,
local muon densities measured for the different thresholds, and
number and energy sum of reconstructed hadrons at the central
detector. As a phenomenological result of KASCADE it should be
remarked that the frequency spectra of all these observables, i.e.
for the different particle components, show a clear kink at the
same integral event rates. This is a strong hint for an
astrophysical source of the knee phenomenon based on pure
experimental data, since the same intensity of the flux
corresponds to equal primary energy. The multi component
measurements of KASCADE, especially the investigation of the
hadronic component, allow the evaluation and improvement of
hadronic interaction models in kinematical and energy ranges not
covered by present collider experiments~\cite{ant99a,ant01}.

The shower sizes $N_e$ and $N^{tr}_\mu$ are used as input
parameters for a neural network analysis to reconstruct the
primary energy on an event by event basis. The necessary 'a
priori' information in the form of probability density
distributions are found by detailed Monte Carlo simulations with
large statistics. The resulting energy spectrum depends on the
high energy interaction model underlying the analyses. But the
spectrum shows a clear kink at $\sim$4-5 PeV and power law
dependences below and above the knee. A simultaneous fit to the
$N_e$ and $N^{tr}_\mu$ size spectra is performed for the
reconstruction of the primary energy spectrum. The kernel function
of this fit contains the size-energy correlations for two primary
masses (proton and iron) obtained by Monte Carlo simulations. This
approach leads to the all particle energy spectrum as a
superposition of the spectra of light and heavy particles. Besides
the use of global parameters like the shower sizes, sets of
different parameters (describing different shower particle
components) are used for neural network and Bayesian decision
analyses for showers with their axes within the central detector
area. Examples of such observables are the number of reconstructed
hadrons in the calorimeter ($E_h>$100 GeV), their reconstructed
energy sum, the energy of the most energetic hadron (`leading
particle' in the EAS), number of muons in the shower center
($E_\mu>$2 GeV), or parameters obtained by a fractal analysis of
the hit pattern of muons and secondaries produced in the passive
calorimeter material. The latter ones are sensitive to the
structure of the shower core which is mass sensitive due to
different shower developments of light and heavy particles in the
atmosphere. Monte Carlo statistics limit the number of parameters
which can be used for one multivariate analysis. Therefore a set
of approaches using different observables are averaged in case of
the actual result of the Bayesian analysis. The resulting
classifications are corrected with misclassification matrices
leading to relative abundances. Afterwards the results are
converted into distributions of the mean logarithmic mass. The
resulting all-particle energy spectrum is shown in
Figure~\ref{espec}~\cite{ant01a}. Figure~\ref{loga-air} shows the
mean logarithmic mass obtained by this approach versus primary
energy~\cite{web99}.

%%%%%%%%%%%%%%%%%%%%%%%%%%%%%%%%%%%%%%
\subsection{CASA-MIA} \label{sec.casamia}

The Chicago Air Shower Array - MIchigan muon Array (CASA-MIA)
detector~\cite{bor94} is an array of surface and underground
plastic scintillators which measure the  electromagnetic and muon
components of air showers. The array is located on Dugway Proving
Grounds, Utah, USA. The CASA-MIA data are used here to determine
the probability, on an event by event basis, that a given data
shower resulted from a `light' or a `heavy' primary. These
distinctions are made by comparison to simulations of proton- or
iron-induced air showers. A composition-independent measure of the
energy is used to search for trends in the composition as a
function of energy.

In order to study composition, a detailed fit of all events is
performed after a series of standard fits determines core
distance, the electron and muon sizes and other shower quantities.
Both the standard and detailed fits are based on the NKG and
Greisen functions; the detailed fits allow more parameters to
remain free.  Three parameters sensitive to composition are
extracted: the density of surface particles $\rho_e$ and the slope
$\alpha$ of the lateral distribution near the core, and the
density of muons $\rho_\mu$ at large core distance. The parameters
$\rho_e$, $\rho_\mu$ and $\alpha$ are tabulated for data and for
simulation events. Data events are classified as `iron-like' or
`proton-like' depending on whether these parameters most resemble
iron or proton simulation events. The K Nearest Neighbor (KNN)
test is employed to quantify this decision~\cite{gla98,gla99c}.
Each data event is placed in the three-parameter space defined by
$\rho_e$, $\rho_\mu$ and $\alpha$. A large set of simulated iron
and proton events also populate the space. Suitably normalized
distances (computed in units of the variance of each parameter,
with correlations included) between the datum and individual
simulation events is calculated. The nuclear types of the five
nearest simulation events are then tallied. The use of K=5 is
optimal for this analysis~\cite{gla98}. Using too few simulation
points is subject to fluctuations, using too many would sample
simulation points with very different primary energies. Each
CASA-MIA event takes one of six possible values, corresponding to
whether it has 0-5 proton neighbors out of the five nearest
neighbors examined. Using separate simulated calibration sets in
place of data, it is found that more than 90\% of events will have
a majority of neighbors of their own species while about 50\% have
all nearest neighbors of their own kind.  The `proton resemblance'
was normalized to the results of analyzing separate simulation
calibration sets of pure iron and of pure protons. For example,
near 100 TeV, the average fraction of proton nearest neighbors for
a pure proton calibration sample is 68\%; at the same energy a
pure iron sample has only 16\%. The average for CASA-MIA data at
this energy is 51\%. The computed proton resemblance is
approximately proportional to $\langle$log(A)$\rangle$, the mean
value of the logarithm of the atomic number of the
sample~\cite{gla98,gla99b}. These data are shown in
Figure~\ref{loga-air}.

%%%%%%%%%%%%%%%%%%%%%%%%%%%%%%%%%%%%%%
\subsection{DICE} \label{sec.dice}

The Dual Imaging Cherenkov Experiment (DICE) is a ground based air
shower detector which is designed to have as little reliance as
possible on the details of the air shower simulations and to have
the capability of comparison with existing direct measurements at
0.1~PeV to provide an assessment of the overall systematic error.
Since the method of imaging hadronic showers in Cherenkov light is
a relatively recent development, we provide some more detailed
description of the DICE detectors and operation. The two DICE
telescopes are located at the CASA-MIA site in Dugway,
Utah~\cite{bor94}). They each consist of a 2m diameter f/1.16
spherical mirror with a focal plane detector of 256 close packed
40~mm hexagonal photomultipliers (PMTs) which provide $\sim 1\deg$
pixels in an overall field of view $16\deg \times 13.5\deg$
centered about the vertical. The telescopes are on fixed mounts
separated by 100~m (see~\cite{boo95,boo97}).

Cosmic-ray events within the field of view produce a focal plane
image at the PMT cluster which corresponds to the direction and
intensity of Cherenkov light coming from the air shower. When the
direction of the air shower core and the distance of the shower
axis from the telescopes are known, simple geometry can be used to
reconstruct the amount of light received from each altitude of the
shower. The amount of Cherenkov light produced is strongly
correlated with the number of electrons in the shower and is used
to estimate the electron size as a function of depth in the
atmosphere from which the location of shower maximum can be
determined. This procedure is essentially geometrical and is
independent of numerical simulations except for calculations which
determine the angular distribution of Cherenkov light around the
shower axis. For each air shower collected a simple time
coincidence is used to identify the same event in both DICE
clusters and CASA-MIA. Further requirements on the correlation of
the DICE images with each other and with the CASA-MIA event
geometry are used to reduce the overall probability of event
mismatches between the detectors to $\sim$10$^{-5}$. The
parameters for each shower are derived from these measured values.
The accuracy of the shower core location derived by CASA is 1-3~m
depending on the overall shower size. The measurement of the
shower arrival direction is accurate to $\sim 0.4\deg$ for larger
showers with some degradation for lower energy events.

Previous work with DICE estimated the shower energy by a simple
translation from the total amount of Cherenkov light in the image
and the geometry of the shower~\cite{bapjl97}. In this
work~\cite{swo00} a more accurate estimate of energy is derived
from a combination of the amount of Cherenkov light and the
location of shower maximum size ($X_{max}$, in atmospheric depth
g/cm$^2$) determined by each DICE telescope. This is desirable
since the lateral distribution and intensity of Cherenkov light at
a given total energy depends both on the primary particle mass,
hence mean $X_{max}$, and the distance of the measurement from the
shower core. A fit for the total shower energy and primary
particle mass is made to the geometry, Cherenkov size and
$X_{max}$ location in the two DICE telescopes. The form of the
Cherenkov size function used in these fits is derived from the
results of simulations using the program CORSIKA 4.50 with the
VENUS interaction model~\cite{corsika98}. The derived $X_{max}$
fitting function has a constant shower elongation rate parameter
and assumes simple superposition for providing primary particle
mass dependence.

\begin{figure}
  \centering
  \includegraphics[height=5.0in]{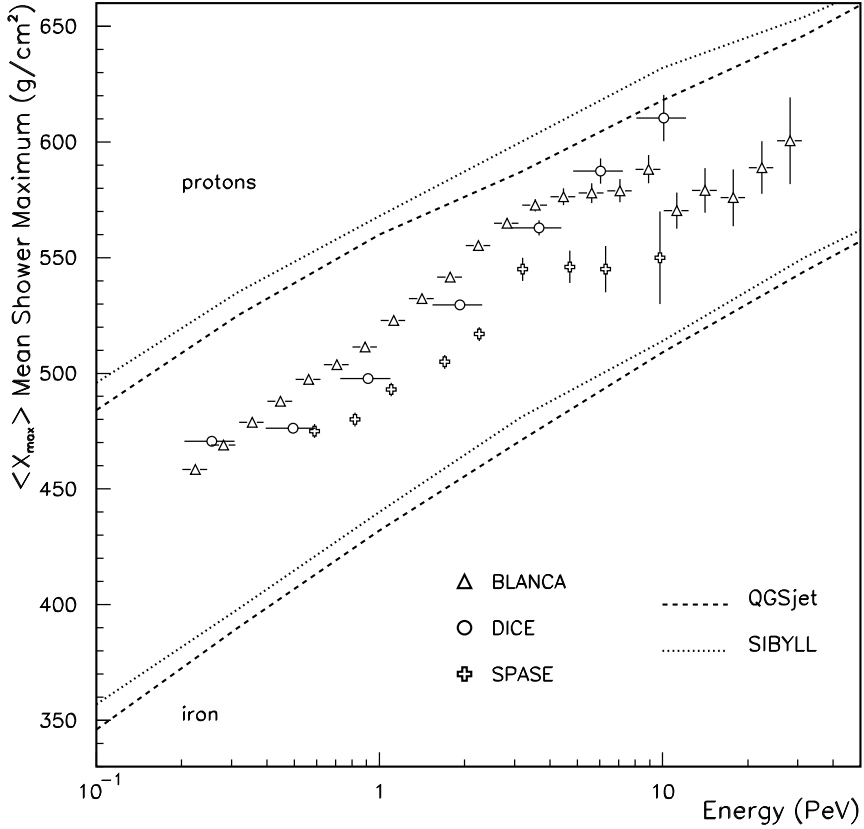}
  \caption{Results from experiments reporting
  $\langle X_{max} \rangle$~\cite{fow01,swo00,dic99}.
  Also shown are the results from model calculations discussed in text.
  The SIBYLL~\cite{sibyll} calculation is from Table~\ref{tab-xmax},
  the QGSjet is from~\cite{pry01}.}
  \label{xmax}
\end{figure}

With the acceptance cuts used during analysis the effective
collection geometry is determined by the instrument Monte Carlo to
be $\sim$3300 m$^2$~sr, making the overall collecting power
$\sim$125,000 m$^2$~sr~days. The results for the mean location of
$X_{max}$ versus energy are shown in Figure~\ref{xmax}.

%%%%%%%%%%%%%%%%%%%%%%%%%%%%%%%%%%%%%%
\subsection{BLANCA} \label{sec.blanca}

The Broad LAteral Non-imaging Cherenkov Array (BLANCA) takes
advantage of the CASA-MIA particle array installation by
augmenting it with 144 angle-integrating Cherenkov detectors.
Located in Dugway, Utah at an atmospheric depth of 870 g
cm$^{-2}$, BLANCA uses the CASA trigger to collect Cherenkov light
and records the Cherenkov lateral distribution from cosmic ray
events in the energy range of the knee. The CASA trigger threshold
imposes an energy threshold of $\sim 100$ TeV on the Cherenkov
array. However, BLANCA analysis uses events with a 200 TeV minimum
to avoid composition bias introduced from the CASA trigger.

Each BLANCA detector contains a large Winston cone~\cite{winston}
which concentrates the light striking an 880\,cm$^2$ entrance
aperture onto a photomultiplier tube.  The concentrator has a
nominal half-angle of 12.5$^\circ$ and truncated length of 60\,cm.
The Winston cones were aligned vertically with $\sim 0.5^\circ$
accuracy.  A two-output preamplifier increases the dynamic range
of the detector.  The minimum detectable density of a typical
BLANCA unit is approximately one blue photon per cm$^{2}$.

BLANCA operated between January 1997 and May 1998. After applying
data quality cuts, about 460 hours of Cherenkov observations
remain corresponding to an exposure of 212,000~m$^2$~sr~days. The
detectors were calibrated on a nightly basis by using the
circularly symmetric property of the cosmic ray air showers. This
method was also verified with a portable LED flasher. The LED
system was then used in the laboratory to produce single
photoelectrons in a PMT with good charge resolution to obtain the
absolute calibration of the BLANCA system. A full description of
the detector can be found in~\cite{fow00}.

The lateral distribution of Cherenkov light is fit to an empirical
functional form to yield the normalization at 120~m ($C_{120}$)
and the exponential slope between 30 and 120~m ($s$). To interpret
these measurements, a large set of air shower simulations were
generated with the CORSIKA Monte Carlo package. To access the
systematics inherent in the choice of interaction models, four
separate hadronic models were used - QGSjet, VENUS, SIBYLL and
HDPM.  Ten thousand events were generated for each hadronic model
in four primary groups - protons, helium, nitrogen and
iron~\cite{fort99b}. The simulated air shower events were then
processed through a full BLANCA detector simulation~\cite{fort99a}
to obtain ``fake'' data. These fake data were used to find the
optimum transfer functions between the observables $C_{120}$ and
$s$, and the energy and mass of the primary particle. The energy
error distribution depends on the primary mass and energy.
Assuming a mixed cosmic ray composition, the energy resolution for
a single air shower is approximately 12\% for a 200 TeV shower,
falling to 8\% for energies above 5 PeV. In a similar way, a
transfer function from slope, $s$, to \Xmax\, was determined and
the results are plotted in Figure~\ref{xmax}. The quantity that is
of direct astrophysical interest is the mean nuclear mass of the
primary. In obtaining an absolute nuclear mass from the Cherenkov
observables we chose to translate directly from $(C_{120},s)$ to
$\lnA$ to avoid compounding the errors that would occur by using
primary energy and \Xmax\, as intermediate quantities. For the
purposes of this paper however, a translation through our \Xmax\,
values has been done for direct comparison with other groups. The
results are shown in Figure~\ref{loga-qgs}. Full details of the
BLANCA analysis are in~\cite{fow01}.

% Something about the energy spectrum and the knee???

%%%%%%%%%%%%%%%%%%%%%%%%%%%%%%%%%%%%%%
\subsection{SPASE/VULCAN/AMANDA} \label{sec.vulcan}

An experiment to measure the electron, muon and air-Cherenkov
components of $\sim$1~PeV air-showers has been established at the
geographic South Pole (688~g/cm$^{2}$).  The experiment comprises:
the SPASE-2 scintillator array~\cite{dic00a}; the VULCAN
air-Cherenkov array~\cite{dic00b}, and the deep under-ice muon
detector AMANDA~\cite{and00}.  Simultaneous measurement of the
electron size, high energy ($>$500~GeV) muon content and lateral
distribution of Cherenkov light should allow a composition
measurement that is relatively insensitive to model assumptions.

SPASE-2 (a $1.6\times10^{4}$~m$^2$ array of plastic scintillator
modules) provides an event trigger with a threshold of
$\sim$50~TeV for proton primaries. The SPASE-2 data are used to
determine the shower core/direction to an rms accuracy of 4
m/1$^\circ$ at 1~PeV. Digitized waveforms from the nine element
VULCAN array (similar in design to AIROBICC~\cite{kar95}) are used
to reconstruct the lateral distribution of Cherenkov light. For a
subset of events the shower axis passes through the AMANDA
detector and the ice-Cherenkov signal can be used to estimate the
$> 500$~GeV muon content of the shower. The aperture of the
SPASE-2/AMANDA-B10 telescope is $\approx$100~m$^{2}$sr.

The lateral distribution of Cherenkov light from air-showers is
closely related to their longitudinal development~\cite{pat83} and
can be used to provide an indirect measurement of the depth of
shower maximum ($X_{max}$) and primary energy ($E$). The MOCCA
air-shower simulation~\cite{hil95} (coupled with the SIBYLL
hadronic interaction model~\cite{sibyll}) has been used to deduce
$X_{max}(E)$ from the Cherenkov data obtained using the VULCAN
array during 1997 and 1998 (see Figure~\ref{xmax}). The systematic
uncertainties in $X_{max}/E$ are $\approx$ 15~g/cm$^{2}$/20\,\%.
This work is described in detail in~\cite{dic99}
and~\cite{dic00b}. A preliminary investigation of the distribution
of $X_{max}$ at around 1~PeV suggests a ratio of
(proton+helium)/all = $0.13^{+0.03}_{-0.02}$ for the
MOCCA/SIBYLL-17 model~\cite{hin98}. However the probability of
compatibility with the data is low (2\%). A much better fit is
obtained by shifting the $X_{max}$ distributions to the mean
values predicted by the CORSIKA/VENUS model~\cite{fzka5828}:
(proton+helium)/all = $0.39^{+0.03}_{-0.01}$ with a probability of
40\%.

For the subset of SPASE-2/VULCAN/AMANDA coincidences, a
measurement of muon content from AMANDA can be combined with the
VULCAN energy determination to estimate primary mass. Preliminary
results from this method are consistent with the SPASE-2/VULCAN
$X_{max}$ result but have limited statistics.  These statistics
will be increased by more than a factor of three by the addition
of data from 1999 and 2000 in which the projected area of the
AMANDA was increased.

In addition the SPASE-2 and VULCAN surface arrays have been useful
calibration and surveying tools for the AMANDA detectors. It is
anticipated that a surface air-shower detector would form a
natural part of any future ice-Cherenkov neutrino observatory.

%%%%%%%%%%%%%%%%%%%%%%%%%%%%%%%%%%%%%%
\section{Simulation Models for Air Showers}\label{sec:sim}

The measurement of extensive air showers (EAS) is presently the
only way to study cosmic rays with high statistics at energies
near or above the knee of the all-particle spectrum. The
properties of the primary cosmic rays have to be deduced from the
shape and the particle content of the EAS.  The energy of the
primary is approximately reflected in the total number of
secondaries produced and its mass can be deduced from the shower
form. An iron primary e.g., consisting of 56 nucleons, would
produce roughly a superposition of 56 nucleon showers of energy
E/56, thus being shorter with a shower maximum higher up in the
atmosphere than a proton shower of the same energy.

Since experiments at knee energies or above cannot be calibrated
with a test beam the interpretation of EAS measurements is
performed by comparison of the experimental data with model
predictions of the shower development in the atmosphere.
Therefore, unfortunately, quantitative results rely on the model
assumptions and on the quality of the simulation of particle
interactions and transport in the atmosphere.

While the electromagnetic interaction (responsible for
electromagnetic showers, ionization, Cherenkov light production,
etc.) and the weak interaction (responsible for decays of unstable
secondaries) are rather well understood, the major uncertainties
in Monte Carlo simulations of EAS arise from the hadronic
interaction models. The development of an EAS for a given energy
and primary is strongly dependent on the inelastic cross-sections
$\sigma_{\rm inel}$ of primary and secondary particles with air
and on the fraction of the available energy transferred into
secondary particles (usually termed {\em inelasticity}). Large
cross-sections and high inelasticity produce short showers, while
small cross-sections and low inelasticity produce long showers
penetrating deep into the atmosphere. Unfortunately a systematic
variation in cross-section and/or inelasticity in the model would
directly translate into uncertainties of the primary mass
estimates obtained from the height of the shower maximum.

The present theoretical understanding, in the form of QCD, cannot
be applied to calculate the hadronic inelastic cross-section or
the particle production from first principles.  Therefore hadronic
interaction models are usually a mixture of basic theoretical
ideas and empirical parameterizations tuned to describe the
experimental data at lower energies. Unfortunately none of the
collider experiments register particles emitted into the extreme
forward direction. These particles, being produced in collisions
with rather small momentum transfer, carry most of the hadronic
energy and are of the greatest importance in the development of
EAS, since they transport a large energy fraction deep into the
atmosphere.  Moreover, since cosmic ray energies exceed those
accessible by accelerators by several orders of magnitude, models
have to be extrapolated well beyond the range of firm knowledge.
Here one has to rely on theoretical guidelines.

%%%%%%%%%%%%%%%%%%%%%%%%%%%%%%%%%%%%%%
\subsection{CORSIKA and the hadronic interaction models}

The most widely used air shower model is the CORSIKA
program~\cite{corsika1,corsika2}.  It is a fully 4-dimensional
Monte Carlo code, simulating in great detail the evolution of EAS
in the atmosphere initiated by photons, nucleons, or nuclei.
Single particles are tracked, accounting for energy loss,
deflection due to multiple scattering and the Earth's magnetic
field, kinematically correct decays of unstable particles, and
electromagnetic and hadronic interactions. CORSIKA contains
various hadronic interaction models which permits a comparison and
an estimate of systematic errors due to model assumptions and
uncertainties in the hadronic interaction.  The available models
are HDPM~\cite{cap,corsika1}, VENUS~\cite{venus},
QGSjet~\cite{qgsjet}, SIBYLL~\cite{sibyll}, and
DPMJET~\cite{dpmjet} and, for the energies below E$_{\rm lab} =
80$ GeV per nucleon, GHEISHA~\cite{gheisha}.

VENUS, QGSjet, and DPMJET model hadronic interactions of nucleons
and nuclei on the basis of the Gribov-Regge theory
(GRT)~\cite{gribov}, which most successfully describes elastic
scattering and, via the optical theorem, the total hadronic
cross-section as a function of energy. In GRT the observed rise of
the cross-sections is a consequence of the exchange of multiple
{\em supercritical pomerons}. Inelastic processes are described by
{\em cut pomerons} leading to two color strings each, which
fragment subsequently to color neutral hadrons. The probability of
n exchanged and m cut pomerons is uniquely given by the theory.
While this is common for all implementations of the GRT, they
differ in the way the production and decay of the strings is
realized. (A good introduction to GRT models is given
in~\cite{venus}.)

SIBYLL is a {\em minijet} model. It simulates a hadronic reaction
as a combination of one underlying soft collision in which two
strings are generated and a number of minijet pairs leading to
additional pairs of strings with higher $p_\perp$ ends. In this
model the rise of the cross-section is due to the minijet
production only.

HDPM is a purely phenomenological generator which uses detailed
parameterizations of $p\bar{p}$ collider data for particle
production. The extension to reactions with nuclei and to energies
beyond the collider energy range are somewhat arbitrary.

GRT models simulate nucleus-nucleus collisions in great detail
allowing for multiple interactions of nucleons from projectile and
target. The configurations of interacting projectile and target
nucleons are determined with a Glauber-type geometrical picture of
the nucleons in a nucleus~\cite{glauber}.  HDPM and SIBYLL employ
for nuclear projectiles the superposition model assuming an
independent reaction for each of the projectile nucleon.

QGSjet, DPMJET, and SIBYLL account for minijet production, which
becomes dominant at higher energies and leads to increasing
cross-sections and more high-$p_\perp$ particles. While in QGSjet
and DPMJET minijets contribute only partially to the rise of the
cross-section, they are solely responsible for it in SIBYLL.

VENUS is the only model taking into account the interactions of
intermediate strings and secondary hadrons with each other,
leading to better agreement with final states as measured in
collider experiments. In DPMJET these interactions are only
implemented for secondary hadrons.

For energies below 80 GeV GHEISHA is used, which was written
around 1985 to simulate the interactions of GeV secondaries from
$e^+e^-$ collisions with typical detector materials. GHEISHA is a
phenomenological model, i.e. it parameterized particle production
to agree with beam test results.

Detailed comparisons of the models available in CORSIKA have been
performed~\cite{fzka5828,habil} revealing differences on the 25\%
to 40\% level and proving some of the models unable to describe
aspects of the experimental results at all (e.g. SIBYLL and
KASCADE results~\cite{kasc_modtest}). In the last three years some
of the models were updated and new models were added to CORSIKA.

QGSjet.00 is an updated version of QGSjet with an increased
Pomeron-nucleon coupling leading mainly to a cross-section with a
5\% larger value at $\sqrt{s} = 1800$ GeV, which is well within
the variation of the experimental results~\cite{cdf,e710,e811} of
about 10\%. This was attempted to explain the trigger and hadron
rates in the KASCADE experiment~\cite{risse}. However, while the
EAS data could be reproduced better by QGSjet.00 the rapidity
distributions for $p-\bar{p}$ collisions were no longer reproduced
by the model. Therefore it was decided to stick with the previous
cross-sections.

An error in the implementation of the DPMJET cross-sections was
discovered and corrected which brought the values down by about
10\%.

A new version of SIBYLL (2.1)~\cite{sibyll2} was produced as a
response to the deficit of muons in SIBYLL 1.6. SIBYLL now allows
for multiple string production in soft collisions, which yields
more muons and modifies the resulting cross-sections. Also better
structure functions have been introduced to improve the
extrapolations to high energies, and the treatment of
nucleus-nucleus collisions was improved. Version 2.1 of SIBYLL is
not yet officially released. The results shown here are based on a
preliminary version which was available to the authors for
evaluation.

The model {\sc neXus}~\cite{nexus} emerges as a new and more
powerful program constructed by the authors of VENUS and QGSjet.
{\sc neXus} is based on GRT but includes also hard interactions
and attempts to describe all types of high energy interactions in
a uniform way. {\sc neXus} version 2 is implemented in the next
CORSIKA release.

URQMD~\cite{urqmd} is an alternative to GHEISHA for low energy
hadron interactions. It has been used in heavy ion collision
experiments. It performs a more detailed modelling and would avoid
problems with GHEISHA at a few GeV energies, but it turns out not
to be free of problems either.

%%%%%%%%%%%%%%%%%%%%%%%%%%%%%%%%%%%%%%
\subsection{Inelastic Cross-Sections}

A summary of experimental and predicted cross-sections is shown in
Figure~\ref{fig-cross}.  Back in 1998 the variation of the p-air
cross-section in experiments as well as in models amounted to
about 25\% at $3\times 10^{15}$ eV, and to about 40\% at $10^{18}$
eV (shown in the left panel). For nucleus-air cross-sections the
model uncertainty was only 10-15\% due to the averaging effect of
many nucleons inside the nucleus. The right panel shows the
situation after some of the model authors have revised their
cross-sections and new theoretical predictions have been
published. HDPM, VENUS and MOCCA are not shown since they are no
longer supported by their respective authors.

\begin{figure}
\begin{center}
\includegraphics[width=0.49\textwidth]{sigmod_p_air_bw_old.epsf}
\includegraphics[width=0.49\textwidth]{sigmod_p_air_bw_new.epsf}
\end{center}
\caption{\small Inelastic p-air cross-sections.  {\em left:}
Situation in 1998 {\em right:} Situation in 2000}
\label{fig-cross}
\end{figure}

Recent calculations of the p-air cross-sections from Frichter et
al.~\cite{frichter} agree well with the QGSjet cross-sections.
However, beyond $10^{17}$ eV the predictions become slightly
higher than those of QGSjet. Block et al. also~\cite{block}
performed new calculations by using a QCD inspired
parameterization of accelerator data.  They simultaneously fit the
total p-p cross-sections, the ratio of real to imaginary part of
the forward scattering amplitude $\rho$, and the nuclear slope
parameter B. Then the p-p cross-sections were converted via
Glauber theory into p-air cross-sections.  Their result agrees
very well with the cross-sections used in the QGSjet model.

Extracting cross-sections directly from EAS measurements is not
possible.  Usually the attenuation length $\Lambda$ of EAS in the
atmosphere is measured which relates to the cross-sections
$\sigma_{\rm inel}^{\rm p-air}$ via $\Lambda = {\rm k}\cdot
14.6~{\rm m}_{\rm p} / \sigma_{\rm inel}^{\rm p-air}$, where k
depends strongly on the average inelasticity (and the inelasticity
distribution) of p and $\pi$ reactions and on the energy.  While
early and naive models used the inelasticity as one of the free
parameters, in GRT models the inelasticity distribution emerges
inevitably as a result of more fundamental properties of the
interaction model.  Usually k is assumed from an EAS model, which
biases the final results for all the experimental cross-sections.
In addition, different groups have been using different values for
k. The experimental results can be easily brought into agreement
with each other and with the theoretical calculations by slightly
modifying k~\cite{block}. For the AGASA and Fly's Eye data such a
correction brought the experimental values in the energy range
$10^{16}$-$10^{18}$ eV down by 12-20\%, which now are in good
agreement with the cross-sections as predicted by Block et al. or
QGSjet.

As a result of various modifications the model predictions at knee
energies agree now to within about 6\%.

%%%%%%%%%%%%%%%%%%%%%%%%%%%%%%%%%%%%%%
\subsection{Shower Development}

\begin{figure}[b]
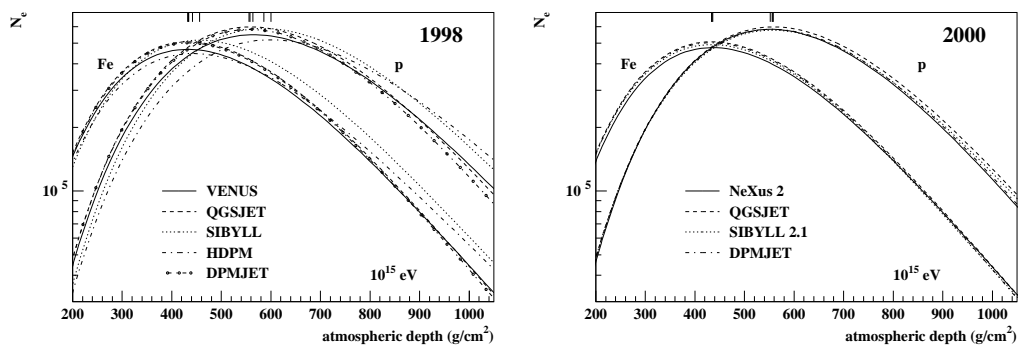

\begin{center}
\includegraphics[width=0.49\textwidth]{longi_bw.epsf}
\includegraphics[width=0.49\textwidth]{longi_bw2.epsf}
\end{center}
\caption{\small Longitudinal shower development. The positions of
the shower maxima are indicated at the upper rim of the picture.
{\em left:} Situation in 1998 {\em right:} Situation in 2000}
\label{fig-long}
\end{figure}

The cross-sections together with the inelasticity determine the
longitudinal development of an EAS, which in turn is closely
related to the most important shower observables: particle number
at ground, height of shower maximum and total energy in the
electromagnetic component.  In Figure~\ref{fig-long} average
shower curves of 10$^{15}$ eV showers from different models are
shown.

While for the old models the position of ${\rm X}_{\rm max}$
varied by about 7.5\% for protons and 5.5\% for iron showers,
these number with the new models are only 1\% and 0.5\%
respectively.  Also the particle numbers at ground level agree now
much better than two years ago.  The variations for p and Fe
showers were reduced from 80\% and 50\% to 14\% and 3\%,
respectively.

Calculations with URQMD instead of GHEISHA yield differences
mainly in the low energy part of the showers. E.g. with URQMD
there are about 30\% more baryons in the energy range 2-10 GeV and
a factor of 2 less for 0.1-2 GeV than with GHEISHA. Also, while
URQMD tends to produce 10\% more muons below 4 GeV and 10\% less
above 4 GeV, the total number of muons stays approximately the
same in both cases.

\begin{figure}
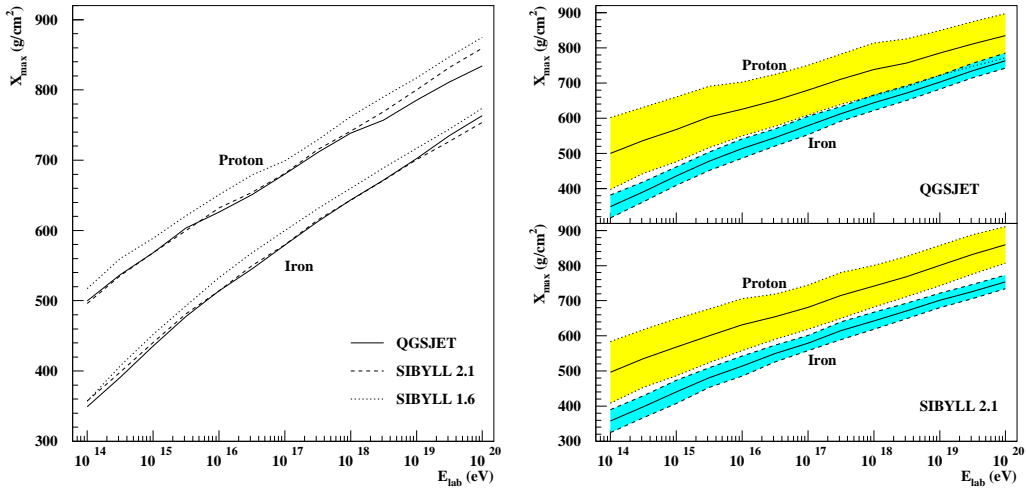

\begin{center}
\includegraphics[width=0.49\textwidth]{xmax.epsf}
\includegraphics[width=0.49\textwidth]{xmax2.epsf}
\end{center}
\caption{\small Height of shower maximum X$_{\rm max}$ as a
function of energy.  {\em left:} The curves for QGSjet, a
preliminary version of SIBYLL 2.1 and SIBYLL 1.6 are shown. {\em
right:} Average values together with the 1-$\sigma$ band
indicating the standard deviation of the X$_{\rm max}$
distribution.~\cite{heck}} \label{fig-xmax}
\end{figure}

The position of the shower maximum X$_{\rm max}$ and its variation
with energy are often used to estimate the mean mass and mass
composition of primary cosmic rays. In Figure~\ref{fig-xmax}
calculated X$_{\rm max}$ values are shown for QGSjet, SIBYLL 1.6
and SIBYLL 2.1.

None of the models show a linear relation as is often assumed in
oversimplified arguments. As compared to SIBYLL 1.6, the new
SIBYLL predicts about 20 g/cm$^2$ smaller X$_{\rm max}$ values for
both proton and iron induced showers. SIBYLL 2.1 and QGSjet agree
very well for iron showers and for proton showers until about
10$^{18}$ eV.  Beyond this energy QGSjet gives smaller values. In
the right panel of Figure~\ref{fig-xmax} fluctuations are also
shown.  The shaded bands indicate the 1-$\sigma$ region around the
average X$_{\rm max}$ values. Note that the fluctuations of
X$_{\rm max}$ for proton showers are of the order of 100 to 60
g/cm$^2$, i.e. not much smaller than the separation between
protons and iron. These fluctuations make particle identification
using X$_{\rm max}$ for single events very difficult and require
high statistics measurements to determine the average values
reliably.  In the case of QGSjet the differences between proton
and iron showers decrease with energy, and consequently the bands
for proton and iron start to overlap at high energies, while for
SIBYLL 2.1 they remain separate. Thus according to QGSjet the
separation of different masses becomes more difficult with energy
than according to SIBYLL 2.1. The X$_{\rm max}$ values are
summarised in Table~\ref{tab-xmax}.
\begin{table}
\begin{center}
\begin{tabular}{|c|cc|cc|}
\hline
 & \multicolumn{2}{|c}{QGSjet} & \multicolumn{2}{|c|}{SIBYLL 2.1} \\
 & proton & iron  & proton & iron \\
\hline showers  &  700 & 200 & 500 & 200 \\ \hline Energy   &
X$_{\rm max}$ &  X$_{\rm max}$ & X$_{\rm max}$ & X$_{\rm max}$ \\
 (eV)    & (g/cm$^2$) & (g/cm$^2$) & (g/cm$^2$) & (g/cm$^2$) \\

\hline \hline $10^{14}$             & 500 $\pm$102 &  349 $\pm$ 33
& 496 $\pm$ 88 & 357 $\pm$ 32  \\ $3.16 \times 10^{14}$ & 537
$\pm$ 94 &  391 $\pm$ 29   & 535 $\pm$ 83 & 398 $\pm$ 31  \\
$10^{15}$             & 568 $\pm$ 92 &  435 $\pm$ 27   & 568 $\pm$
81 & 440 $\pm$ 33  \\ $3.16 \times 10^{15}$ & 603 $\pm$ 87 &  478
$\pm$ 26   & 600 $\pm$ 76 & 481 $\pm$ 28  \\ $10^{16}$ & 626 $\pm$
77 &  514 $\pm$ 28   & 632 $\pm$ 74 & 514 $\pm$ 28  \\ $3.16
\times 10^{16}$ & 651 $\pm$ 74 &  545 $\pm$ 24   & 654 $\pm$ 64 &
550 $\pm$ 24  \\ $10^{17}$             & 680 $\pm$ 71 &  579 $\pm$
26   & 681 $\pm$ 63 & 580 $\pm$ 22  \\ $3.16 \times 10^{17}$ & 711
$\pm$ 71 &  612 $\pm$ 21   & 715 $\pm$ 65 & 615 $\pm$ 25  \\
$10^{18}$             & 739 $\pm$ 75 &  644 $\pm$ 22   & 742 $\pm$
59 & 643 $\pm$ 25  \\ $3.16 \times 10^{18}$ & 757 $\pm$ 68 &  672
$\pm$ 21   & 769 $\pm$ 57 & 671 $\pm$ 22  \\ $10^{19}$ & 786 $\pm$
64 &  702 $\pm$ 20   & 800 $\pm$ 57 & 701 $\pm$ 21  \\ $3.16
\times 10^{19}$ & 811 $\pm$ 63 &  735 $\pm$ 21   & 832 $\pm$ 56 &
726 $\pm$ 20  \\ $10^{20}$             & 834 $\pm$ 63 &  764 $\pm$
22   & 859 $\pm$ 52 & 753 $\pm$ 19  \\ \hline
\end{tabular}
\end{center}
\caption{\small Average X$_{\rm max}$ and standard deviation for
proton and iron showers and QGSjet and SIBYLL 2.1 models. The
values are determined from 200 showers per energy point for iron
and from 500 to 700 showers per energy for protons by fitting the
shower maximum to each individual shower and averaging over the
fit results~\cite{heck}.} \label{tab-xmax}
\end{table}

%%%%%%%%%%%%%%%%%%%%%%%%%%%%%%%%%%%%%%
\subsection{Collider Results}

Another, and sometimes underestimated, source of uncertainties are
the errors of the collider results which are used to tune the
interaction models. There are a few instances where the inherent
precision of those results is revealed.

The different inelastic p-$\bar{\rm p}$ cross-sections as measured
by experiments at Fermilab were already
mentioned~\cite{cdf,e710,e811}. Their results vary from $80.0 \pm
2.2$ mb to $71.7 \pm 2.0$ mb at $\sqrt{s} = 1800$ GeV or ${\rm
E}_{\rm lab}=1.7\times 10^{15}$ eV.  This corresponds to an 11\%
uncertainty already below the knee, and clearly much more for the
highest energy cosmic rays.

\begin{figure}
\begin{center}
\epsfig{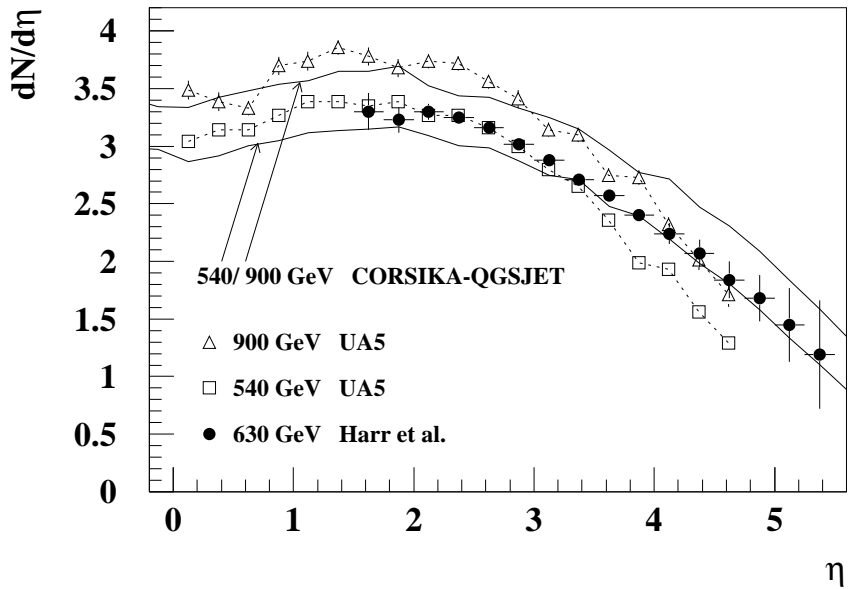}
\end{center}
\caption{\small Pseudorapidity distributions as measured and
simulated.} \label{fig-harr}
\end{figure}

A second example is the pseudorapidity distribution of secondary
particles produced in p-$\bar{\rm p}$ collisions. The UA5
measurements from $\sqrt{s}=200$ to 900 GeV~\cite{ua5} have been
widely used to tune hadronic interaction models. GRT models always
had severe difficulties reproducing the rapidity densities
dN/d$\eta$ near $\eta = 0$ and the slope in dN/d$\eta$ at around
$\eta =4$ at the same time.  However, more recent measurements by
Harr et al.~\cite{harr} show a clearly flatter distribution than
the UA5 ones, that fit very well to the GRT predictions (see
Figure~\ref{fig-harr}). In the near-forward region ($\eta \approx
4$) the differences in the particle densities amount to about
25\%.

%%%%%%%%%%%%%%%%%%%%%%%%%%%%%%%%%%%%%%
\subsection{Summary of air shower models}

In the last two years much has changed in EAS simulations.  There
is a clear trend of convergence between different hadronic
interaction models, which, to some extent, is due to objective
improvement of the physics put into the models. Presently QGSjet
seems to be the best model showing a good overall agreement with a
variety of experimental results. Typical differences in
cross-sections or inelasiticities with other upcoming models
(SIBYLL 2, {\sc neXus}) are of the order of 10\%. Unfortunately
the number of independent models may be effectively reduced from 5
to 2 ({\sc neXus} and SIBYLL 2), which will make it more difficult
to estimate systematic uncertainties in the future, although a
completely new version of DPMJET is in development\cite{roesler}.
Further improvement will be difficult since there is a variety of
uncertainties at the 10\% level in collider results, in
theoretical asumptions and, even larger ones, in experimental air
shower results, that enter into construction of EAS models. This
makes it rather unlikely that EAS model predictions will get much
better in the near future. A promising source of new information
on hadronic interactions in the longer term will be the
experiments at LHC and the new heavy ion colliders, even though
CERN decided not to pursue the proposed experiments for very small
angle secondaries.

It is also unfortunate that many of the experiments investigating
the composition and spectra at the knee have ceased operation
(CASA-MIA-BLANCA, DICE, EAS-TOP, HEGRA-AIROBICC). Also here it
will be in future more difficult to combine different aspects of
the problem and to estimate the systematics. However, high
statistics and good quality data are a good basis for substantial
further improvement, perhaps with more model independent analysis
techniques.

%%%%%%%%%%%%%%%%%%%%%%%%%%%%%%%%%%%%%%%%%%%%%%%%%%%%%%%%%%%%%%%%%%%%%%%%%%%%%%%
\section{Discussion of Results} \label{sec.disc}

%%%%%%%%%%%%%%%%%%%%%%%%%%%%%%%%%%%%%%
\subsection{Energy Spectra}
\begin{figure}
  \centering
  \includegraphics[height=5.0in]{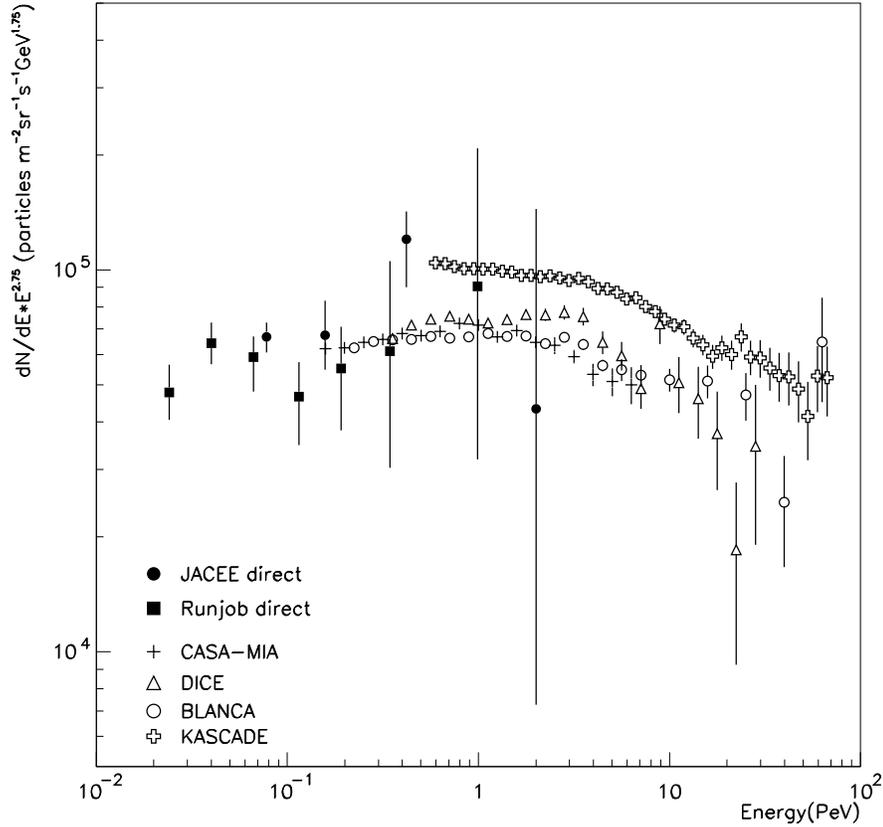}
  \caption{All-particle energy spectra measurements near the `knee' from
various experiments presented at the workshop}
  \label{espec}
\end{figure}

Part of the intent of the workshop, and this report, is to examine
how the various experiments are consistent in the measurement of
composition in the region of the knee. As a precursor to this
comparison it is useful to see the measurements from the various
experiments of the energy spectrum of cosmic rays in this region.
These give some information on the systematic differences which
may be present between the experiments, since the establishment of
an absolute energy scale and the geometrical factor (m$^2$sr s) of
an experiment is often the most demanding task in the overall
analysis effort. Figure~\ref{espec} shows a comparison of the
measured all-particle energy spectra from both direct and indirect
measurements; only statistical errors are shown. The vertical
intensity axis has been multiplied by $E^{2.75}$, to emphasize
differences from the overall spectrum of particles. The knee is
clearly visible around 3~PeV in the air shower measurements. The
overall flux normalization between the direct and indirect
measurements is well within the errors, which are large and
dominated by the limited geometrical factor of the direct
measurements. Any of these measurements typically has a systematic
error in the energy scale which is on the order 10-20\%, this
translates to a 20-40\% shift in the overall intensity values
normalization on this plot. The air shower measurements seem to
agree almost too well with this level of expected systematic
error.

%Something about the recommended function for knee determination:
% f(E) = E^{\gamma_1}(1+(E/E_knee)^\epsilon)^(\gamma_1-\gamma_2/\epsilon)

%%%%%%%%%%%%%%%%%%%%%%%%%%%%%%%%%%%%%%
\subsection{Composition estimates}

The derivation of composition from measurements such as those of
mean shower maximum $\langle X_{max} \rangle$, or derived from the
electron and muon numbers, need a choice of the simulation to be
used. As shown in Figure~\ref{xmax}, both SIBYLL and QGSjet
predict an overall logarithmic increase in the depth of shower
maximum with energy. Here, to evaluate the size of systematic
effects in the simulations, the SIBYLL data are taken from
Table~\ref{tab-xmax}~\cite{heck} and the QGSjet data are from the
recent work of Pryke~\cite{pry01} who has calculated similar
quantities independently. The slope of this increase is fairly
similar, but the overall normalization differences can have an
impact on the mass estimates.

\begin{figure}
  \centering
  \includegraphics[height=5.0in]{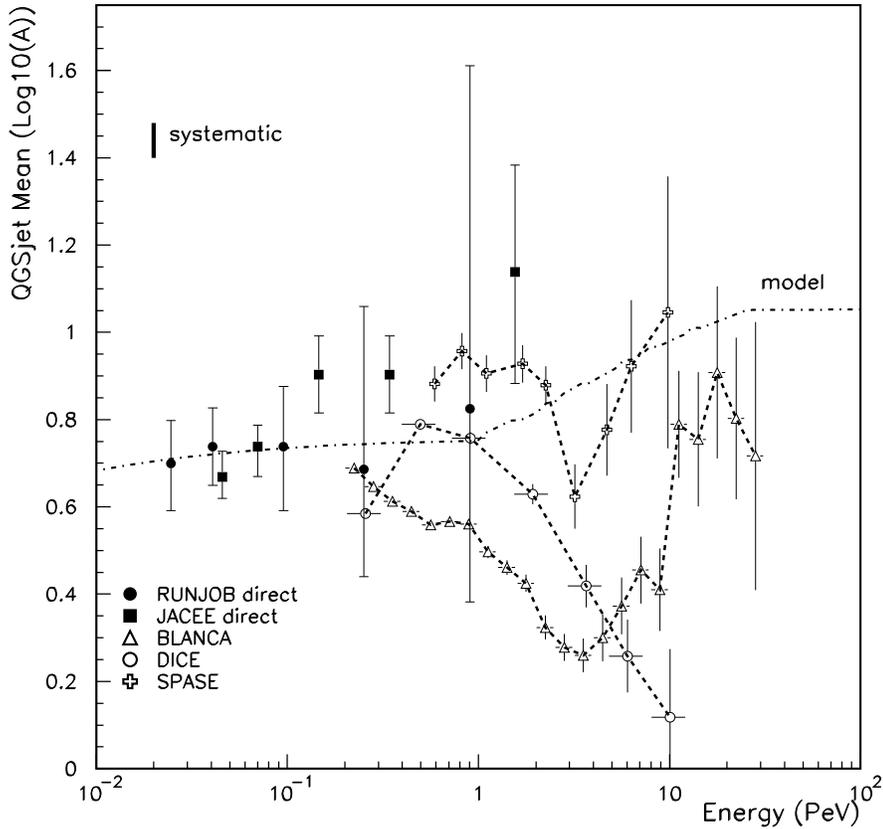}
  \caption{Results on shower maximum analyzed with the QGSjet model}
  \label{loga-qgs}
\end{figure}

As an example we can derive mean logarithmic mass composition from
Figure~\ref{xmax} by assuming the mass curves for other nuclei can
be derived from a simple logarithmic interpolation between the
proton and iron curves shown here. (This means that at any given
energy $\langle log(A) \rangle= log(56)\times(X_{max}^{p}-\langle
X_{max}\rangle )/(X_{max}^{p}-X_{max}^{Fe}$), where $X_{max}^{p}$
and $X_{max}^{Fe}$ are the model predictions and $\langle
X_{max}\rangle$ is the experimentally measured value). Although
this procedure is really only accurate if simple superposition is
valid, it can give us an approximate basis on which to compare
$\langle log(A) \rangle$ results between experiments. The results
of this procedure for the $\langle X_{max} \rangle$ results of
Figure~\ref{xmax} using QGSjet are shown in Figure~\ref{loga-qgs}
where $\langle log10(A) \rangle$ is shown versus energy, with the
various data sets connected by dashed lines. Also shown here are
the results from the direct measurements of the same quantity.
There is good agreement between the direct/indirect measurements,
although the direct measurements have poor statistics in the knee
region.

\begin{figure}
  \centering
  \includegraphics[height=5.0in]{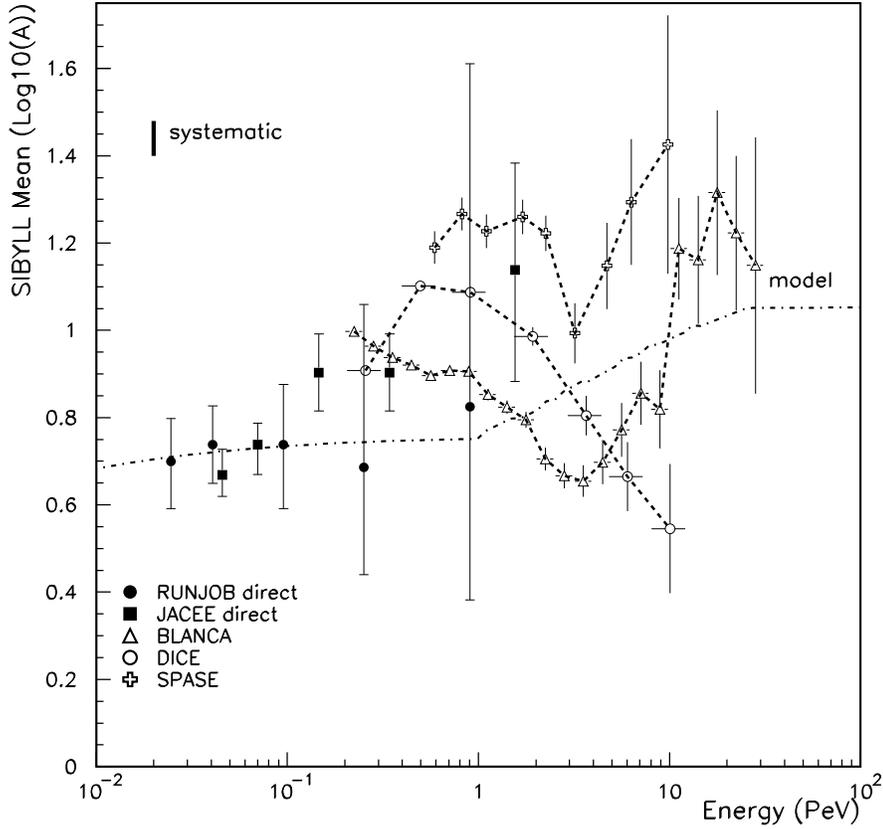}
  \caption{Results on shower maximum analyzed with the SIBYLL model}
  \label{loga-sib}
\end{figure}

To evaluate the systematic errors in mass estimate, a similar
procedure can be made with the SIBYLL model. The results of this
are shown in Figure~\ref{loga-sib}. Here the mean masses are
larger, corresponding to the generally deeper shower maxima
expected for this model shown in Figure~\ref{xmax}. However the
general structure of the composition changes is preserved and
again these are not inconsistent with the directly measured
values.

Within these Cherenkov air shower measurements, all compositions
show a decrease in mean mass from around 1PeV to 3PeV and two
experiments (BLANCA and SPASE-VULCAN) show an increase in mean
mass above 3~PeV. A systematic shift in the energy scale of any of
these air shower experiments will directly affect the inferred
mass. The bar labelled `systematic' at the left hand side of
Figures~\ref{loga-qgs} and~\ref{loga-sib} shows the shift in
$\langle log(A) \rangle$ expected for a $20$\% shift in energy
scale. This is smaller than the systematic shift introduced
between an anaylsis based on QGSjet or SIBYLL 2.1, the two curves
shown in Figure~\ref{xmax}.

The goal of composition measurements near the knee is to look for
apparent changes in the makeup of the all-particle spectrum which
can reveal the origin of this feature. The most basic test which
can be made is to examine whether the change in slope at the knee
occurs at the same magnetic rigidity for all nuclear species. This
would be expected if this feature has an origin either in an
intrinsic steepening of the source spectrum of cosmic rays,
possibly because diffusive shock acceleration becomes weaker, or
because the particles begin to escape from the Galaxy more easily
at these energies. An identification of the knee with a simple
break in magnetic rigidity would significantly reduce the range of
possible origins for this feature.

The signature of a constant magnetic rigidity spectral bend is
relatively easy to evaluate in terms of $\langle log10(A)
\rangle$. Since the all-particle spectrum of cosmic rays measures
particle intensities at similar total particle energies ($E=\gamma
Amc^2$, $A$=atomic mass, $m$=nucleon mass), at the same magnetic
rigidity ($R=\gamma mcA/Z$, since $\beta\approx 1$ at these
energies) different nuclear species should show a spectral bend at
a total energy $E=ZRc$, where $Z$ is the nuclear charge. For
example, if a bend occurs for all nuclei at $R$=1PV this will be
observed at $E$=1~PeV for protons and at $E$=26~PeV for iron
nuclei. The heavier nuclei are expected to bend at higher values
of $E$, meaning that the mean mass should steadily increase across
the knee region.

An estimate of the size of this effect can be made by applying a
simple model extrapolating from measured cosmic ray elemental
abundances at lower energies. An extension of a model of the
Galaxy in which particles have an exponential distribution of
pathlengths and an energy dependent leakage probability can be
easily made into the `knee' region~\cite{swordy_model}. These
models produce an equilibrium intensity of particles in the Galaxy
which are the result of subjecting a source rigidity spectrum to
losses by either spallation in the interstellar medium or escape
from the Galactic confinement volume. Introducing a spectral break
from $\propto R^{-2.7}$ to $\propto R^{-3.0}$ at 1PV produces a
increase in $\langle log10(A) \rangle$ shown by the dot-dash curve
labeled model in Figure~\ref{loga-qgs} and Figure~\ref{loga-sib}.

This is a relatively robust prediction since the only assumption
made is that all nuclear species have the same source magnetic
rigidity spectra. The source elemental abundances have been fit to
lower energy data~\cite{swordy_model}. Any model which produces a
bend in the rigidity spectra of this size will produce a similar
increase in $\langle log10(A) \rangle$. The simple model does not
agree well with either of the sets of derived mean compositions
from the data. The measured values are consistently lighter than
expected in two of the experiments. The Cherenkov experiments show
a significant tendency to become lighter into the knee region at
3~PeV for all hadronic models. This does not agree with a simple
rigidity bend origin for the knee.

\begin{figure}
  \centering
  \includegraphics[height=5.0in]{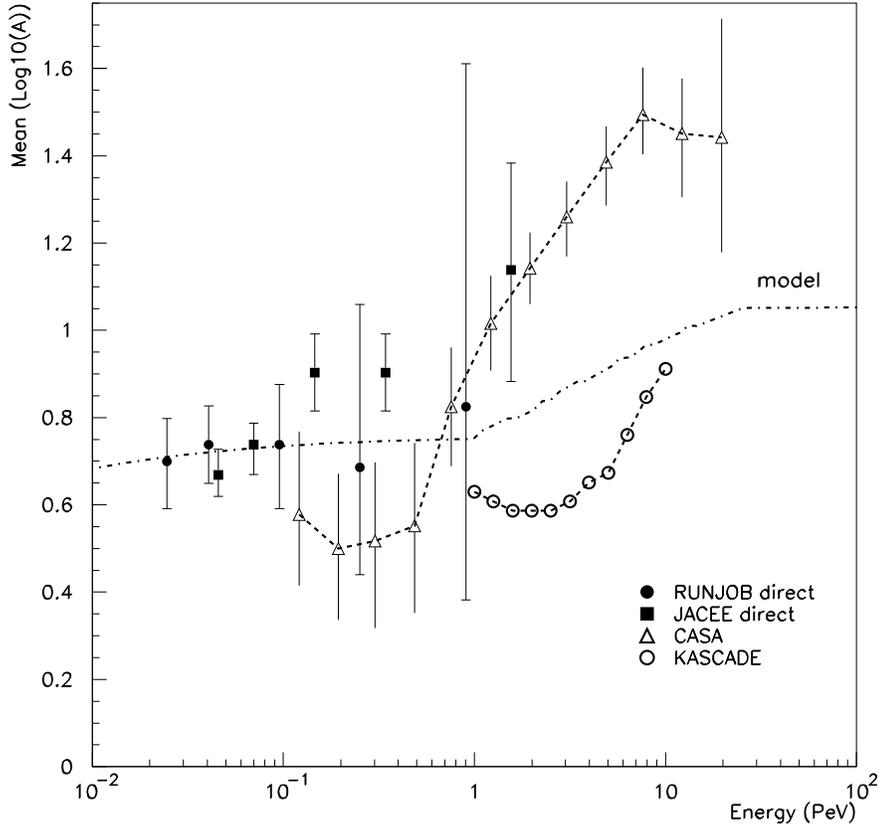}
  \caption{Mass measurements from KASCADE and CASA-MIA compared
with direct measurements.}
  \label{loga-air}
\end{figure}

Figure~\ref{loga-air} shows the estimates of $\langle log10(A)
\rangle$ derived from CASA-MIA using MOCCA/SIBYLL1.6 and KASCADE
using CORSIKA/QGSjet. Also shown are the results from the direct
measurements. The simple rigidity bend model is the dot-dashed
curve discussed above. The derived compositions agree in general
with the direct measurement normalization. The KASCADE data show a
trend towards greater mean mass even though there is a dip towards
lighter values around 3~PeV similar to that seen with the
Cherenkov experiments in Figure~\ref{loga-qgs}. CASA-MIA provides
the most significant shift to heavy composition of all the
experiments discussed here.

%%%%%%%%%%%%%%%%%%%%%%%%%%%%%%%%%%%%%%%%%%%%%%%%%%%%%%%%%%%%%%%%%%%%%%%%%%%%%%%
\subsection{Multi-Species Fits and Multi-Parameter Correlations}
\label{sec.multi}

The discussion in the last section is in terms of the mean value
of a parameter, be it $\langle log10(A) \rangle$ or $\langle
X_{max} \rangle$. In fact, although fluctuations are severe, even
a simple air shower experiment provides much more information in
the form of the distributions and correlations of the observable
quantities. This information can be used to improve the quality of
the results --- for instance providing a fractional composition
breakdown into several elemental groups versus energy, rather than
simply a mean mass. However, and probably more importantly given
the highly indirect nature of the EAS experiments, the information
also provides a means to check in detail the consistency of the
data with simulations, and hence to search for problems in the
shower and detector Monte Carlos.

In the past the limited statistics of the data often meant that
distribution analyses were not attempted --- this is no longer the
case. Additionally full shower and detector simulations, complete
with all relevant fluctuations, were not carried out --- this is
now a standard part of any serious experiment. These two
conditions being fullfilled it becomes possible to compare the
measured distribution of a composition sensitive parameter to the
expected distributions for a range of primary masses. We normally
assume that the range of possible primary particles in the knee
region is bounded by protons at the light end and iron nuclei at
the heavy. If this is the case the measured distribution must be
reproducible by a sum of simulated distributions for proton, iron
and some number of intermediate atomic masses.

It is sometimes claimed that the severe inherent fluctuations of
the air shower process limit the number of mass groups that may be
fitted. In fact only the statistics of the data and the accuracy
of the simulations impose such limits --- in principle an
arbitrary number of components can be deconvolved. In practice the
impact of potential systematic errors in the simulations must be
carefully assessed.

If the observed distribution cannot be fit by an appropriate sum
of simulated components then either the detector simulation is not
adequately reproducing the measurement process, or the shower
simulation is flawed. If the shower simulation is suspected then,
as discussed in Section~\ref{sec:sim}, by far the most likely
culprit is the high energy hadronic interaction model. In this way
it can become possible to both choose an appropriate interaction
model, {\it and} measure the primary composition simultaneously.

\begin{figure}
\begin{center}
\includegraphics[width=0.49\textwidth]{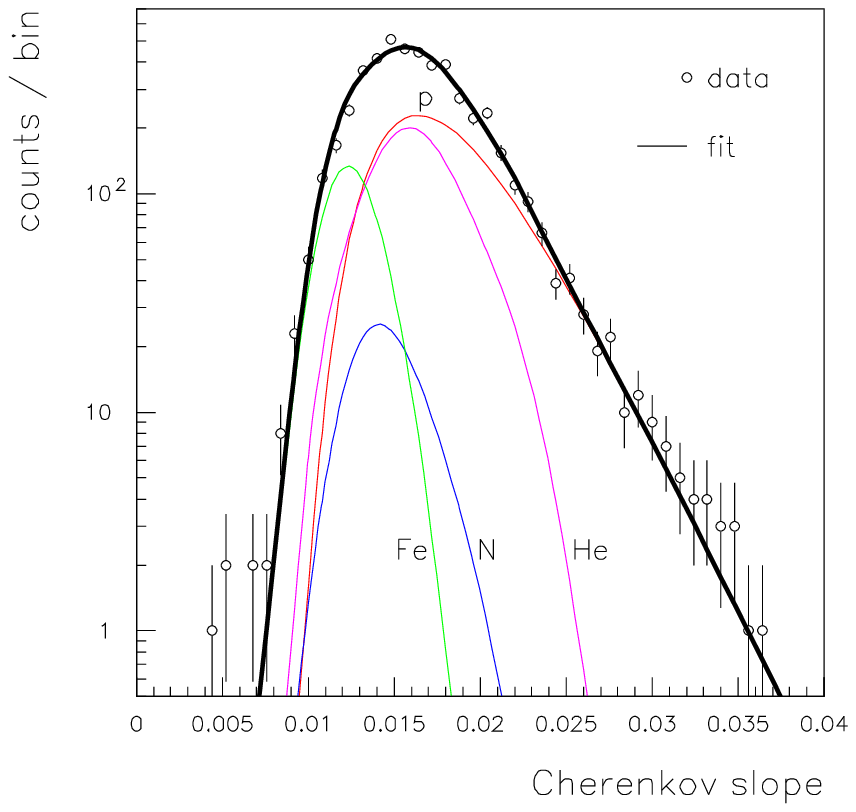}
\includegraphics[width=0.49\textwidth]{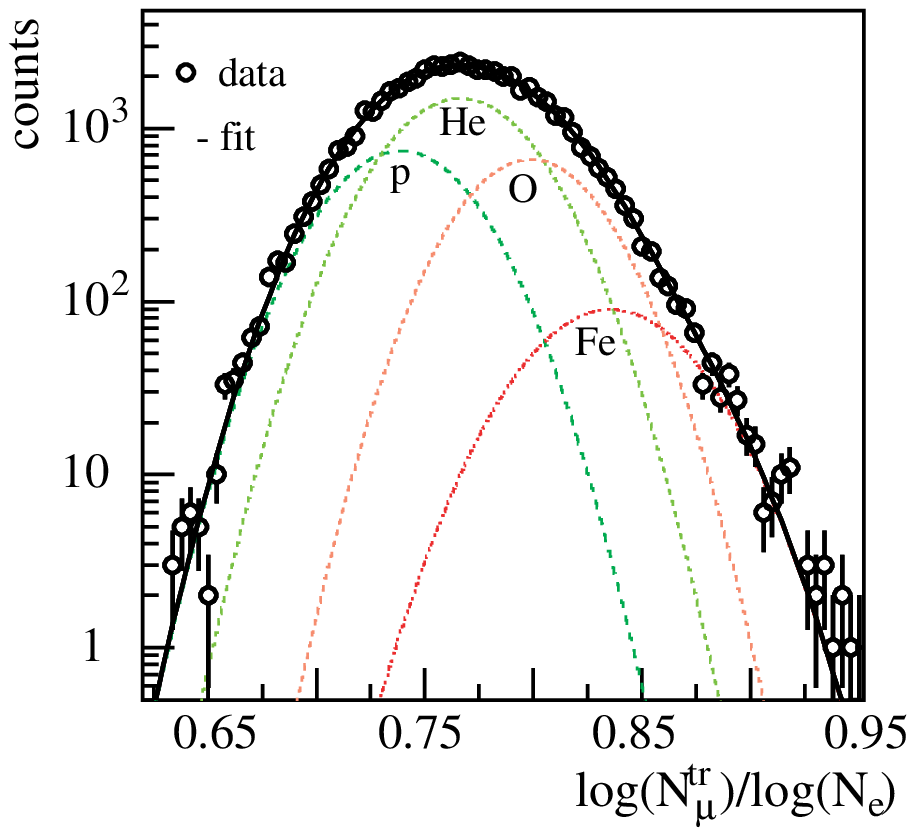}
\caption{Two examples of multi-species fits to air shower
observables; BLANCA\cite{fow01} at left, KASCADE\cite{kam00} at
right. See text for details.} \label{fig:multicomp}
\end{center}
\end{figure}

Two examples of sucessful multi species fits are shown in
Figure~\ref{fig:multicomp}. Both plots show the distribution of a
shower observable which is stongly correlated with primary mass
for a group of events in a band of primary energy around 3~PeV. At
left we have the distribution of the exponential slope of the
Cherenkov lateral distribution from BLANCA, and at right the ratio
of the truncated muon number to electron number from KASCADE. The
binned data is fit to a sum of the expected distributions for four
primary species as simulated by CORSIKA-QGSjet, and after
processing through the detector simulation, and event
reconstruction procedures. The vertical axis is logarithmic in
both cases to emphasize the tails of the distribution where
inadequate simulations are most likely to be revealed. We can see
that both experiments can be well fit by a four species model, and
that neither appears to demand a sub-proton or trans-iron
component. In fact both data sets demand a four species fit ---
the measured distribution {\it cannot} be adequately modeled with
fewer (e.g.\ a simple proton/iron combination). It must be
emphasized that such plots are a recent development, and a major
step forward in air shower studies.

A trend in recent experiments is to measure multiple parameters of
the air shower on an event-by-event basis. It has sometimes been
claimed that this can improve the resolution with which primary
mass can be measured. However shower fluctuations are so severe
that they easily dominate over instrumental resolution for any
given parameter in a well designed experiment. Additionally the
fluctuations are highly correlated between shower observables ---
in most cases what one is effectively measuring is the fluctuation
of the elasticity (and multiplicity) of the first interaction. The
real benefit of multi-parameter measurements is that not only the
distributions of the indvidual observables, but also their cross
correlations, can be used to compare with simulations.

\begin{figure}
\begin{center}
\includegraphics[width=\textwidth]{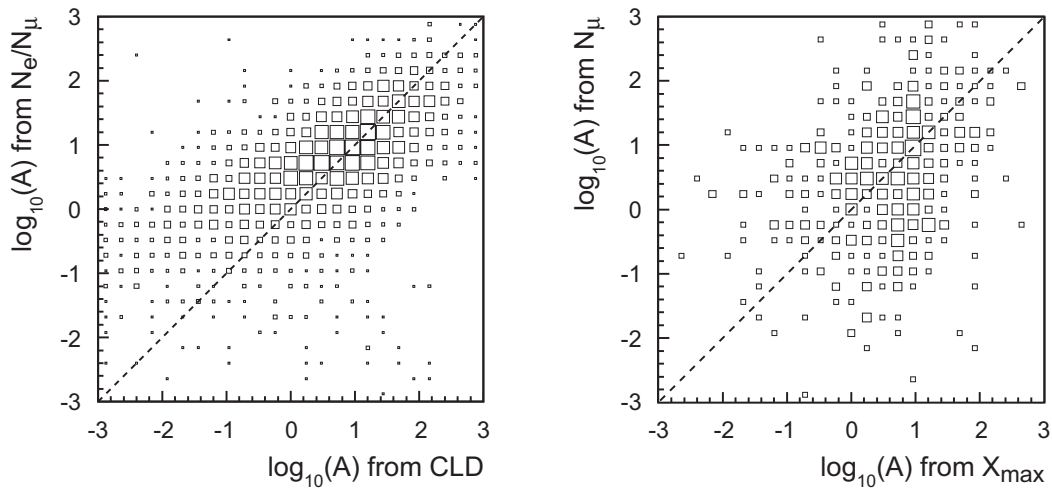}
\caption{Two examples of event-by-event correlations between mass
estimates near 3 PeV derived from different air shower
observables; BLANCA\cite{fow01} at left from Cherenkov Lateral
Distribution and Electron Muon sizes, DICE\cite{swo00} at right
from the Cherenkov image and Muon sizes. See text for details.}
\label{fig:multipar}
\end{center}
\end{figure}

Figure~\ref{fig:multipar} shows two examples from the BLANCA and
DICE extensions to the CASA-MIA system (see
sections~\ref{sec.blanca} and~\ref{sec.dice}). BLANCA and DICE
respectively measure the lateral and longitudinal distributions of
Cherenkov light; CASA and MIA simultaneously measure the muon and
electron sizes. Both plots are for a band of primary energy around
3~PeV (energy being estimated from the Cherenkov observables). The
density of points in the $xy$ plane is shown after making simple
parametric transformations from observables to the logarithm of
the primary mass~\cite{swo00,fow01}. Simulation plots are very
similar for a mixed primary composition. The appearance of points
on the figure with severly non-physical atomic masses serves to
illustrate that the fluctuations of the air shower process are as
large as the difference between the extreme ends of the mass scale
which we are trying to measure. Note that the apparently better
correlations in the case of BLANCA is at least partly due to the
use of $N_e/N_\mu$ rather than $N_\mu$ alone as the mass estimator
--- most of what is observed is that events with flat Cherenkov
lateral distributions have larger electron sizes at ground level.

%%%%%%%%%%%%%%%%%%%%%%%%%%%%%%%%%%%%%%%%%%%%%%%%%%%%%%%%%%%%%%%%%%%%%%%%%%%%%%%
\section{Conclusions} \label{sec.conclusions}

The experimental situation in measurements of cosmic-ray
composition near the knee of the spectrum has benefited in recent
years from several new, relatively large, experiments. In the past
these investigations were often undertaken with a few detectors
and the simulation power for air showers was limited. Modern
experiments have many more channels of information and simulation
codes have become extensive and sophisticated. In general the
experiments discussed here do not suffer from statistical errors
of sampling the shower, many measurements are taken for each
shower. However the fact remains that trying to determine the
identity of the nucleus which initiates an air shower is a very
difficult task. A fundamental limit for this task seems to be
connected to the accuracy of the air shower models which are used
to derive the composition from the data. At present the systematic
errors in estimating the absolute energy scale and shifts between
the air shower models limit the accuracy in $\langle log10(A)
\rangle$ to $\sim 0.1$. This is to be compared with an expected
shift of $\sim$0.25 for a simple rigidity bend. There are mixed
results from the experiments reported here for the existence of a
simple bend. While the ground array measurements show a tendency
for a heavier composition with increasing energy, the air
cherenkov experiments show a tendency to lighten into the knee
region with some increase in mean mass beyond 3 PeV.

Several actions could improve this situation in the future. The
accuracy of the modelling techniques can be improved if better
accelerator results are available in the future as discussed
previously. This also includes good measurements for small angle
secondaries which at present do not exist. The direct measurements
could be improved by longer exposures in space of high energy
cosmic ray experiments. In particular this will help with the
normalization of systematic errors in energy scale and mean mass
measurements with air showers. Plans for using the Space Station
for large payloads, such as the ACCESS mission which was discussed
at the workshop (see http://access.uchicago.edu), can improve the
statistics near 1PeV by about a factor of 50 over those published
here. There are also other possible techniques to measure samples
of cosmic rays at high energies. One of these discussed at the
meeting was the prospect for measuring the direct emission of
Cherenkov light produced by the particle before it interacts in
the atmosphere using imaging techniques~\cite{kie01}. This could
be a way to get significantly better elemental resolution at high
energies with good statistics. Another promising approach is the
multi component unfolding analysis discussed in Section
\ref{sec.multi}. The KASCADE experiment expects to extend this
technique\footnote{Since this workshop this work has been
discussed at the 27th International Cosmic Ray Conference in
Hamburg, Germany, 2001. Please see the conference proceedings for
details\cite{som01}.} to energies of $10^{18}$eV.

The workshop would like to thank the Adler Planetarium in Chicago
for all their help with organizing this meeting and providing an
excellent venue for this event. Additional information discussed
at the workshop but not directly presented here can be found at
http://knee.uchicago.edu

%%%%%%%%%%%%%%%%%%%%%%%%%%%%%%%%%%%%%%%%%%%%%%%%%%%%%%%%%%%%%%%%%%%%%%%%%%%%%%%


\begin{thebibliography}{99}

\bibitem{ces83} P. O. Lagage and C. J. Cesarsky, Astron.\ Astrophys.,
{\bf 118}, 223 (1983) and {\bf 125}, 249 (1983).

\bibitem{erl97} A.D. Erlykin and A.W. Wolfendale, J. Phys. G, v23,
(1997), p979.

\bibitem{jul72} E. Juliusson, P. Meyer, and D. M\"uller, Phys.
Rev. Lett. 29 (1972) 447.

\bibitem{arq00} F. Arqueros et al., Astron. Astrophys. 359 (2000)
682.

\bibitem{agl99} M. Aglietta et al., Astrop. Phys. 10 (1999) 1.

\bibitem{som01} P. Sommers rapporteur report, Proc. 27th Int.
Cosmic Ray Conf. (Hamburg) 2001.

\bibitem{wil95} R.J. Wilkes et al., Proc 24th Int. Cosmic Ray
Conf. (Rome) 2, (1995) 697.

\bibitem{bur86} T.H. Burnett et al., Phys. Rev. Lett., 51, (1986) 1010.

\bibitem{bur90} T.H. Burnett et al., Ap.J. 349, (1990), 25.

\bibitem{asa91} K. Asakamori et al., Proc. 22nd Int. Cosmic Ray
Conf. (Dublin), 2, (1991), 97.

\bibitem{asa93} K. Asakamori et al., Proc. 23rd Int. Cosmic Ray
Conf. (Calgary), 2, (1993), 21 \& 25.

\bibitem{ols95} E.D. Olsen, PhD thesis, (1995), University of
Washington.

\bibitem{asa98} K. Asakamori et al., Ap.J. 502 (1998) 278.

\bibitem{cherry99} M. L. Cherry et al. Proc 26th Int. Cosmic Ray
Conf. (Salt Lake City), 3, (1999), 187.

\bibitem{apa99} A.V. Apananseko et al., Proc. 26th Int. Cosmic Ray
Conf. (Salt Lake City), 3 (1999) 163.

\bibitem{apa01} A.V. Apanaseko et al., Astrop. Phys., in
press.


\bibitem{hot80} N. Hotta et al., Phys. Rev. D, 22 (1980) 1.

\bibitem{oka87} M. Okamoto and T. Shibata, Nucl. Instr. and Meth.,
A257 (1987) 155.

\bibitem{fuj89} T. Fujinaga et al., Nucl. Instr. and Meth., A276
(1989) 317.

\bibitem{apa99a} A.V. Apananseko et al., Proc. 26th Int. Cosmic Ray
Conf. (Salt Lake City), 3 (1999) 231.

\bibitem{ich93} M. Ichimura et al., Phys. Rev. D, 48 (1993) 1949.


\bibitem{kla97} H. O. Klages et al., Nucl. Phys. B Proc. Suppl.
52B (1997) 92.

\bibitem{eng99} J. Engler et al., Nucl. Instr. Meth., A427 (1999)
528.

\bibitem{ant99} T. Antoni et al., Proc. 26th Int. Cosmic Ray
Conf. (Salt Lake City), KASCADE collaboration, 15 contributions.

\bibitem{kam99} K.-H. Kampert, FZKA report 6345, Forschungszentrum
Karlsruhe 1999.

\bibitem{ant99a} T. Antoni et al., J. Phys. G: Nucl. Part. Phys. 25 (1999) 2161

\bibitem{ant01} T. Antoni et al., accepted by J. Phys. G: Nucl. Part. Phys.
2001.


\bibitem{web99} J.H. Weber et al., Proc. 26th Cosmic Ray Conf., 1 (1999) 341

\bibitem{ant01a} T. Antoni et al., accepted by Astrop. Phys, 2001.

\bibitem{bor94} A. Borione et al., Nucl. Instr. Meth., A346 (1994)
329.

\bibitem{gla98} M.A.K. Glasmacher, PhD. thesis (1998) University
of Michigan.

\bibitem{gla99c} M.A.K. Glasmacher et al., Astrop. Phys. 12 (1999)
1.
\bibitem{gla99b} M.A.K. Glasmacher et al., Astrop. Phys. 10 (1999)
291.

\bibitem{boo95} Boothby, K., et al. {\em Proc. 24th Int. Cosmic Ray Conference
(Rome)}, 1995, 2, 697.

\bibitem{boo97} Boothby, K., et al. {\em Nucl. Phys. B (Proc. suppl.)} 52B,
166.

\bibitem{bapjl97} Boothby, K., et al. {\em Ap. J. Lett.}, 491, L35-L38, (1997)

\bibitem{swo00} S.P. Swordy and D.B. Kieda Astrop. Phys. 13 (2000)
137.

\bibitem{corsika98} Heck, D., et al. {\em CORSIKA: A Monte Carlo program to
Simulate Extensive Air Showers}, Forschungszentrum Karlsruhe
Report FZKA 6019 (1998)

\bibitem{fow01} J.W. Fowler et al. Astrop. Phys. 15 (2001) 49.

\bibitem{dic99b} J.E. Dickinson and G.M. Spiczak, Nucl. Instr.
Meth. A440, 95.

\bibitem{sibyll}
R.S. Fletcher, T.K. Gaisser, P. Lipari, T. Stanev, {\it Phys. Rev.
D} {\bf 50} (1994) 5710 \\ J. Engel, T.K. Gaisser, P. Lipari, T.
Stanev, {\it Phys. Rev. D} {\bf 46} (1992) 5013

\bibitem{pry01} C. Pryke, Astrop. Phys., 14 (2001) 319.

\bibitem{winston} W. T. Welford and R. Winston, \emph{High Collection
Non-imaging Optics}, Academic Press, San Diego, 1989.

\bibitem{fow00} J. W. Fowler, PhD Thesis, University of Chicago, 2000.

\bibitem{fort99b} L. F. Fortson \etal, \emph{Proc.\ 26th Int.\
Cosmic Ray Conf., Salt Lake City}, {\bf 5}, 336 (1999).

\bibitem{fort99a} L. F. Fortson \etal, \emph{Proc.\ 26th Int.\
Cosmic Ray Conf., Salt Lake City}, {\bf 5}, 332 (1999).

\bibitem{dic00a} Dickinson, J. E. et al. {\it Nucl. Inst. and Meth. A} {\bf
440}, 114 (2000).

\bibitem{dic00b} Dickinson, J. E. et al. {\it Nucl. Inst. and Meth. A} {\bf
440},  95 (2000).

\bibitem{and00} Andres E. et al., Astroparticle Phys. {\bf 13} 1 (2000).

\bibitem{kar95} Karle, A., et al. {\em Astroparticle Phys.} {\bf 3} 321 (1995).


\bibitem{pat83} J. R. Patterson and A. M. Hillas, {\it J. Phys. G}, {\bf 9},
1433 (1983).

\bibitem{hil95} A.M. Hillas, {\it Proc. 24th Int. Cosmic Ray Conf.
(Rome)} {\bf 1} 270 (1995).

\bibitem{dic99} Dickinson, J. E. et al. {\it Proc. 26th Int. Cosmic Ray Conf.
(Salt Lake City)\/}, {\bf 3}, 136 (1999).

\bibitem{hin98} J.A. Hinton, {\it PhD. thesis, University of Leeds} (1998).


\bibitem{fzka5828}
J. Knapp, D. Heck, G. Schatz,  {\it Comparison of Hadronic
Interaction Models used in Air Shower Simulations,
Forschungszentrum Karlsruhe} {\bf FZKA 5828} (1996)

\bibitem{corsika1}
D. Heck et al., {\it CORSIKA: A Monte Carlo Code to Simulate
Extensive Air Showers, Forschungszentrum Karlsruhe} {\bf FZKA
6019} (1998)

\bibitem{corsika2}
J. Knapp, D. Heck, {\it Extensive Air Shower Simulation with
CORSIKA: A User's Manual,
 Kernforschungszentrum Karlsruhe} {\bf KfK 5196 B} (1993); for an up to date
version see
 http://www-ik3.fzk.de/\~{}heck/corsika/

\bibitem{cap}
J.N. Capdevielle, {\it J. Phys. G: Nucl. Part. Phys.} {\bf 15}
(1989) 909

\bibitem{venus}
K. Werner, {\it Phys. Rep.} {\bf 232} (1993) 87

\bibitem{qgsjet}
N.N. Kalmykov, S.S. Ostapchenko, {\it Yad. Fiz.} {\bf 56} (1993)
105 \\ N.N. Kalmykov, S.S. Ostapchenko, {\it Phys. At. Nucl.} {\bf
56} (3) (1993) 346 \\ N.N. Kalmykov, S.S. Ostapchenko, A.I.
Pavlov, {\it Bull. Russ. Acad. Sci. (Physics)} {\bf 58} (1994)
1966

\bibitem{dpmjet}
J. Ranft, {\it Phys. Rev. D} {\bf 51} (1995) 64

\bibitem{gheisha}
H. Fesefeldt, {\it The Simulation of Hadronic Showers -Physics and
Application-,\\ Rheinisch-Westf\"alische Technische Hochschule,
Aachen} {\bf PITHA 85/02} (1985)

\bibitem{gribov}
V.N. Gribov, {\it Zh. Eksp. Teor. Fiz.} {\bf 53} (1967) 654
(translated in V. N. Gribov, {\it Sov. Phys. JETP} {\bf 26} (1968)
414 )

%\bibitem{pdg}
%C. Caso et al., (Particle Data Group), {\it Review of Particle
%Physics, Eur. Phys. J.} {\bf C3} (1998) 1

\bibitem{glauber}
R.J. Glauber, G. Matthiae, {\it Nucl. Phys. B} {\bf 21} (1970) 135

\bibitem{habil}
J. Knapp,  {\it Forschungszentrum Karlsruhe} {\bf FZKA 5970}
(1997)

\bibitem{kasc_modtest}
T. Antoni et al., (KASCADE Collaboration), {\it J. Phys. G: Nucl.
Part. Phys.} {\bf 25} (1999) 2161

\bibitem{cdf}
F. Abe et al., (CDF Collaboration), {\it Phys. Rev. D} {\bf 50}
(1994) 5550

\bibitem{e710}
N.A. Amos et al., (E710 Collaboration), {\it Phys. Rev. Lett} {\bf
68} (1992) 2433

\bibitem{e811}
C. Avila et al., (E811 Collaboration), {\it Phys. Lett. B} {\bf
445} (1999) 419

\bibitem{risse}
M. Risse, {\it PhD Thesis University Karlsruhe,
 Forschungszentrum Karlsruhe} {\bf FZKA 6493} (2000)

\bibitem{sibyll2}
R. Engel, T.K. Gaisser, T. Stanev, Proc. 26$^{th}$ ICRC, Salt Lake
City, {\bf 1} (1999) 415\\ R. Engel et al. (in preparation)

\bibitem{nexus}
H.J. Drescher, M. Hladik, S. Ostapchenko, K. Werner,
hep-ph/9903296

\bibitem{urqmd}
M. Bleicher et al., {\it J. Phys. G: Nucl. Part. Phys.} {\bf 25}
(1999) 1859

\bibitem{frichter}
G.M. Frichter et al., {\it Phys. Rev. D} {\bf 50} (1997) 3135

\bibitem{block}
M.M. Block, F. Halzen, T. Stanev, hep-ph/9908222

\bibitem{heck}
D. Heck, {\it private communication} (2000)

\bibitem{ua5}
G.J. Alner et al. (UA5 Collaboration), {\it Z. Phys. C} {\bf 33}
(1986) 1

\bibitem{harr}
R. Harr et al., {\it Phys. Lett. B} {\bf 401} (1997) 176

\bibitem{roesler}
S. Roesler, R Engel, and J. Ranft, {\it Proc. of Int. Conf. on
Advanced Monte Carlo for Radiation Physics, Particle Transport
Simulation and Applications (MC 2000), Lisbon} (2000)
(hep-ph/0012252)

\bibitem{kam00} K-H Kampert et al., 30th International Symposium
on Multiparticle Dynamics, Lake Balaton, Hungary, (2000)

\bibitem{swordy_model} S. Swordy, \emph{Proc.\ 24th Int.\ Cosmic Ray Conf.,
Rome}, {\bf 2}, 697 (1995).

\bibitem{kie01} D.B. Kieda, S.P. Swordy and S.P. Wakely Astrop,
Phys., 15 (2001) 287.

\end{thebibliography}
\end{document}